\documentclass[fleqn,usenatbib]{mnras} 

\usepackage{newtxtext,newtxmath}

\usepackage[T1]{fontenc}
\usepackage{ae,aecompl}
\usepackage{newtxtext,newtxmath}

\usepackage{braket}
\usepackage{graphicx}	
\usepackage{amsmath}	
\usepackage{amssymb}	
\usepackage{hyperref}
\usepackage{amsmath}
\usepackage{graphicx}
\usepackage{comment}
\usepackage{bm}
\usepackage{wasysym}
\usepackage{multirow}
\usepackage{rotating}
\usepackage{cite}
\usepackage{braket}
\usepackage{float}
\let\oldtextbf=\textbf
\renewcommand\textbf[1]{{\boldmath\oldtextbf{#1}}}

\title[DF2 and DF4 are in conflict with $\Lambda$CDM]{The ultra-diffuse dwarf galaxies NGC 1052-DF2 and 1052-DF4 are in conflict with standard cosmology} 
\author[M. Haslbauer et al.]
{Moritz Haslbauer$^{1}$\thanks{E-mail:  \mbox{mhaslbauer@astro.uni-bonn.de} (MH)},
Indranil Banik$^{1}$, Pavel Kroupa$^{1,2}$ and Konstantin Grishunin$^{3}$\\
$^{1}$Helmholtz-Institut f\"ur Strahlen- und Kernphysik, University of Bonn, Nussallee 14-16, D-53115 Bonn, Germany\\
$^{2}$Faculty of Mathematics and Physics, Astronomical Institute, Charles University, V Hole\v{s}ovi\v{c}k\'ach 2, CZ-18000 Praha 8, Czech Republic\\
$^{3}$ Saint Petersburg State University, Universitetskii pr. 28, Saint Petersburg 198504, Russia \\
}
\date{Accepted 2019 August 13. Received 2019 July 3; in original form 2019 April 5}

\pubyear{2019}

\pubyear{2019}
\begin{document}
	\label{firstpage}
	\pagerange{\pageref{firstpage}--\pageref{lastpage}}
	\maketitle
	
	\begin{abstract}
		Recently van Dokkum et al. (2018b) reported that the galaxy NGC 1052-DF2 (DF2) lacks dark matter if located at $20$~Mpc from Earth. In contrast, DF2 is a dark-matter-dominated dwarf galaxy with a normal globular cluster population if it has a much shorter distance near $10$~Mpc. However, DF2 then has a high peculiar velocity wrt. the cosmic microwave background of $886$~$\rm{km\,s^{-1}}$, which differs from that of the Local Group (LG) velocity vector by $1298$~$\rm{km\,s^{-1}}$ with an angle of $117 \, ^{\circ}$. 
		Taking into account the dynamical $M/L$~ratio, the stellar mass, half-light radius, peculiar velocity, motion relative to the LG, and the luminosities of the globular clusters, we show that the probability of finding DF2-like galaxies in the lambda cold dark matter ($\Lambda$CDM) TNG100-1 simulation is at most $1.0\times10^{-4}$ at $11.5$~Mpc and is $4.8\times10^{-7}$ at $20.0$~Mpc. At $11.5$~Mpc, the peculiar velocity is in significant tension with the TNG100-1, TNG300-1, and Millennium simulations, but occurs naturally in a Milgromian cosmology. At $20.0$~Mpc, the unusual globular cluster population would challenge any cosmological model. Estimating that precise measurements of the internal velocity dispersion, stellar mass, and distance exist for $100$ galaxies, DF2 is in $2.6\sigma$ ($11.5$~Mpc) and $4.1\sigma$ ($20.0$~Mpc) tension with standard cosmology. Adopting the former distance for DF2 and assuming that NGC 1052-DF4 is at $20.0$~Mpc, the existence of both is in tension at $\geq4.8\sigma$ with the $\Lambda$CDM model. If both galaxies are at $20.0$~Mpc the $\Lambda$CDM cosmology has to be rejected by $\geq5.8\sigma$.
	\end{abstract}
	
	
	\section{Introduction} \label{sec:Introdcution}

	In the standard hierarchical bottom-up cosmological model, dwarf galaxies void of dark matter are formed in the overdensities of tidal tails and arms during interactions of massive galaxies. The high velocity dispersion of dark matter particles and the relatively shallow gravitational potential compared to their host galaxy prevent tidal dwarf galaxies (TDGs) to capture a significant amount of dark matter \citep{Barnes_1992, Bournaud_2006_456_481,Wetzstein_2007, Bournaud_2008b,Bournaud_2008a,Fouquet_2012,Yang_2014,Ploeckinger_2018,Haslbauer_2019}. 
	
	The Dual Dwarf Theorem states that in any viable cosmological model, interactions between galaxies imply that both primordial and TDGs must exist. In particular, the lambda cold dark matter ($\Lambda$CDM) cosmological model predicts both dark-matter-dominated dwarf galaxies and dark-matter-lacking TDGs with different observed stellar half-mass radii \citep{Kroupa_2010, Kroupa_2012, Dabringhausen_2013,Haslbauer_2019}. The former are primordial galaxies formed by the condensation of gas in dark matter halos. 
	Dark-matter-dominated and dark-matter-poor dwarf galaxies (i.e. TDGs) are distinguishable in a radius--mass plane, because TDGs are formed naked without the help of dark matter and have to be therefore more compact in order to remain gravitationally bound \citep{Kroupa_2012, Dabringhausen_2013,Haslbauer_2019}. 
	
	\citet{Ploeckinger_2018} demonstrated that TDGs can be found in the self-consistent cosmological EAGLE simulation and the study of TDG formation in the Illustris-1 simulation by \citet{Haslbauer_2019} contains a very detailed analysis of their properties as well as of resolution issues.
	
	\citet{vDokkum_2018a} reported that the ultra-diffuse dwarf galaxy (UDG) NGC 1052-DF2 (hereafter DF2\footnote{For the historical record of when this galaxy was discovered see \citet{Trujillo_2019}.}) lacks dark matter based on velocity measurements of $10$ globular clusters (GCs). They derive a velocity dispersion of $\sigma_{\mathrm{intr}}
	<10.5 \, \rm{km \, s^{-1}}$ with $90 \, \%$ confidence corresponding to a total mass of $M_{\mathrm{total}} <3.4 \times 10^{8} \, \rm{M_{\odot}}$ within a radius of $7.6 \, \rm{kpc}$, a stellar mass of $M_{\mathrm*} = 2 \times 10^{8} \, \rm{M_{\odot}}$ by assuming a mass-to-light ratio of $M/L_{\mathrm{V}} \approx 2 \, \rm{\Upsilon_{\odot}}$, and an effective radius along the major axis of $R_{\mathrm{e}} = 2.2 \, \rm{kpc}$ \citep{vDokkum_2018a} corresponding to a $3$D stellar half-light radius of\footnote{The effective radius along the semimajor axis of DF2 can be converted to a 3D circularized half-mass radius by $R_{\mathrm{1/2}} \approx \frac{4}{3} R_{\mathrm{e}} \sqrt{b/a}$, where $b/a$ is the axis ratio being $0.85$ for DF2 \citep{Wolf_2010, vDokkum_2018a}.} $R_{1/2} = 2.7 \, \rm{kpc}$. 
	
	Such an observation of a dark matter deficient galaxy would support that dark matter is indeed separable from dwarf galaxies and would therewith be consistent with the $\Lambda$CDM paradigm if it is a TDG. \citet{Nusser_2019} showed that tidal stripping of the outer dark matter halo of a primordial dwarf galaxy caused by the close galaxy NGC 1052 might also account for the low velocity dispersion of DF2. Such a process would be included in the hydrodynamical simulations.
	
	Nevertheless, the intrinsic velocity dispersion of DF2 is rather uncertain. \citet{Martin_2018} calculate a $90 \, \%$ upper limit of $\sigma_{\mathrm{intr}} < 17.3 \,\rm{km \, s^{-1}}$ implying $M/L_{\mathrm
	{V}} < 8.1 \, \rm{\Upsilon_{\odot}}$, which is comparable with the mass-to-light ratio of dwarf galaxies in the Local Group (LG) and allows a significant amount of dark matter \citep{Martin_2018}. In a following paper \citet{vDokkum_2018b} revised the velocity of the globular cluster GC-98 and derived a new intrinsic velocity dispersion of $\sigma_{\mathrm{intr}}<12.4 \, \rm{km \, s^{-1}}$ at $90 \, \%$ confidence concluding that `[it] is impossible to say whether the galaxy has no dark matter at all but it is clearly extremely dark matter deficient' \citep{vDokkum_2018b}. A very low intrinsic velocity dispersion of around $\sigma_{\mathrm{intr}} = 7.8_{-2.2}^{+5.2} \, \rm{km \, s^{-1}}$ \citep{vDokkum_2018b} of an isolated dwarf galaxy would be, according to 
	\citet{vDokkum_2018a}, in major conflict with modified gravity theories such as Milgromian dynamics \citep[MOND,][]{Milgrom_1983}. However, DF2 is close to the massive elliptical galaxy NGC 1052 and taking into account its external field (which is an important prediction of MOND), DF2 becomes well consistent with MOND \citep{2018MNRAS.480..473F,Kroupa2018DoesTG,Haghi_2019,Mueller_2019}. In fact, DF2 is extremely gas-depleted and could be therefore a satellite of NGC 1052, having lost its gas reservoir via ram-pressure stripping and tidal forces triggered by several close encounters \citep{Chowdhury_2019, Sardone_2019}. 
	
	Based on the surface brightness fluctuation (SBF) method, a distance for DF2 of $D = 19.0 \pm 1.7 \, \rm{Mpc}$ from Earth has been derived \citep{vDokkum_2018c}. However, within the integrated galaxy-wide initial mass function (IGIMF) theory \citep{Kroupa_2003,Yan_2017,Jerabkova_2018}, the IMF depends on the star formation rate (SFR). Therefore, dwarf galaxies that have a low SFR will have a top-light IGIMF compared to massive galaxies \citep{Watts_2018}. This would result in a lower number of giant stars, possibly making DF2 appear to be further away than it actually is \citep{Zonoozi_2019}. Even without the IGIMF theory \citet{Trujillo_2019} obtained a much smaller distance of $D = 14.7 \pm 1.7 \, \rm{Mpc}$ by using the SBF method.
	
	The apparent lack of dark matter and the unusually high luminosity of GC-like objects surrounding DF2 only hold if this dwarf galaxy is indeed located at a distance of approximately $D=20 \, \rm{Mpc}$ from Earth. 
	A revised distance of about $D = 10 \, \rm{Mpc}$ would make DF2 a usual dark-matter-dominated dwarf galaxy with $M/L_{\mathrm{V}}=16.2  \, \rm{\Upsilon_{\odot}}$ \citep[assuming the velocity dispersion derived in][]{Martin_2018} and a GC population comparable to that of other galaxies \citep{Haghi_2019,Trujillo_2019}. 
	However, at such a small distance DF2 has a large radial peculiar velocity of about $v_{\mathrm{pec}}=886 \, \rm{km \, s^{-1}}$ with respect to (wrt.) the cosmic microwave background (CMB) and a velocity of $v_{\mathrm{rel}}=1298 \, \rm{km \, s^{-1}}$ relative to the motion of the LG. Such a high velocity is highly unlikely in the standard $\Lambda$CDM cosmological model which is based on Newtonian dynamics. A possible tension with the $\Lambda$CDM paradigm and a qualitative comparison to the predicted velocity field in MOND cosmological simulations conducted by \citet{Candlish_2016} are investigated inter alia in this work (see Sections~\ref{sec:Results_cosmological_simulations} and \ref{sec:Velocity field of subhaloes}). In the real Universe it is supposed that most of the peculiar velocity of the LG \citep[$v_{\mathrm{pec,LG}} = 627 \pm 22 \, \rm{km \, s^{-1}}$ wrt. the CMB rest frame,][]{Kogut_1993} arises from a putative Great Attractor (GA) being about $62 \, \rm{Mpc}$ away from us \citep{Bell_1988}. Within this distance from the Earth we should thus not expect large directional differences between galaxies that have large peculiar velocities, these being presumably generated by the GA. 
	
	The exact internal velocity dispersion and thus the dark matter content of DF2 remains controversial. Using the Multi Unit Spectroscopic Explorer (MUSE) at the VLT two ESO teams performed the first spectroscopic measurement of the stellar body of DF2 and discovered planetary nebulae (PNe) for the first time in an UDG \citep{Emsellem_2018,Fensch_2018}. Using the Jeans model, \citet{Emsellem_2018} obtain $M/L_{\mathrm{V}}$ ratios between $3.5-3.9(\pm1.8)  \, \rm{\Upsilon_{\odot}}$, which is close to the $2 \sigma$ upper limit of the results by \citet{Martin_2018}. 
	
	In contrast to that, \citet{Danieli_2019} measured the stellar kinematics of DF2 with the Keck Cosmic Web Imager and derived, assuming $D=20 \, \rm{Mpc}$, $M_{*}=(1.0 \pm 0.2) \times 10^{8} \, \rm{M_{\odot}}$, a dynamical mass of $M_{\mathrm{dyn}} = (1.3 \pm 0.8) \times 10^{8} \, \rm{M_{\odot}}$ within the 3D half-light radius of $R_{1/2} = 2.7 \, \rm{kpc}$. In such a situation DF2 would indeed be an outlier compared to dwarf galaxies in the LG \citep[see also fig.~5 in][]{Danieli_2019}. 
	
	\citet{Trujillo_2019} concluded that DF2 has a revised distance of $13.0 \pm 0.4 \, \rm{Mpc}$ based on five redshift-independent distance indicators including the tip of the red giant branch (TRGB, $D =13.4 \, \pm 1.1 \, \rm{Mpc}$) and the SBF method ($D =14.7 \, \pm 1.7 \, \rm{Mpc}$), which were obtained with data from the Hubble space telescope (HST). They also estimated the distances using the luminosities and sizes of its GCs, and a comparison of the stellar luminosity function with that of the galaxy DDO44, which is similar to DF2. They obtain $R_{\mathrm{e}} = 1.4 \pm 0.1 \, \rm{kpc}$, $M_{*} = 6 \times 10^{7} \, \rm{M_{\odot}}$, $M_{\mathrm{tot}} \ga 10^{9} \, \rm{M_{\odot}}$, and $M_{\mathrm{halo}}/M_{*} > 20$ such that DF2 would be consistent with dwarf galaxies of the LG \citep[see fig.~29 in][]{Trujillo_2019}.
	
	\citet{vDokkum_2019} report the discovery of a second dark matter lacking galaxy, NGC 1052-DF4 (DF4), with an intrinsic velocity dispersion of $\sigma_{\mathrm{intr}} = 4.2_{-2.2}^{+4.4} \, \rm{km \, \, s^{-1}}$ located close to NGC 1052 and NGC 1042 in sky projection. Assuming a distance of $20 \, \rm{Mpc}$ this galaxy has a mass within the outermost GC of $M_{\mathrm{TME}}(R<7 \, \rm{kpc}) = 0.4_{-0.3}^{+1.2} \times 10^{8} \, \rm{M_{\odot}}$ estimated by the tracer mass estimator method, $M_{*} = (1.5 \pm 0.4 ) \times 10^{8} \, \rm{M_{\odot}}$, $b/a = 0.89$, $R_{\mathrm{e}} = 1.6 \, \rm{kpc}$, and a GC population similar to that in DF2 (i.e. all seven identified GCs of DF4 have absolute magnitudes $M_{\mathrm{V, 606}} \leq -8.6 \, \rm{mag}$).

	The unusual properties of DF2 (and of DF4) and its observed peculiar velocity constrain theoretical cosmology which must account for its existence. In this work we take into account the dynamical M/L ratio, structural property, mass, peculiar velocity, motion relative to the LG, and the GC population of these dwarfs. We concentrate mostly on DF2 because it has better observational constraints at present than DF4. By using simulations of the Illustris \citep{Nelson_2015Illustris}, ``The Next Generation'' (TNG) Illustris \citep{Nelson_2018TNGrelease}, and EAGLE \citep{schaye2014eagle} projects, we assess the existence of a DF2-like galaxy in the $\Lambda$CDM framework over a wide range of distances. 
	
	The structure of the paper is as follows: In Section~\ref{sec:Methods} we briefly introduce the Illustris, IllustrisTNG, and EAGLE simulations. The probabilities of detecting a DF2-like galaxy in these simulation runs and a study of its GC population are given in Section~\ref{sec:Results}. In Section~\ref{sec:Discussion} we discuss our results and their implication for the $\Lambda$CDM paradigm including a comparison of the peculiar velocity field in IllustrisTNG and Millennium $\Lambda$CDM simulations and in MOND. This is followed by a conclusion in Section~\ref{sec:Conclusion}. Throughout this work we adopt for observed quantities a Hubble constant of $H_{0} = 70 \, \rm{km \, s^{-1} \, Mpc^{-1}}$ at the present time. 
	
	\section{Methods} 
	\label{sec:Methods}
	In this section we briefly introduce the Illustris-1, IllustrisTNG, and EAGLE cosmological simulations and describe the selection criteria and search-algorithm for simulated DF2- and DF4-like dwarf galaxies. In Section~\ref{sec:Velocity field of subhaloes} we also employ the much larger ($685 \, \rm{cMpc}$ side) box of the Millennium simulation to quantify robustly the peculiar velocity field in a $\Lambda$CDM cosmology.
	
	In the Illustris, IllustrisTNG, and EAGLE projects, comprehensive physical models are implemented allowing a detail study of galaxy formation and evolution in a self-consistent context across cosmic time. We stress that the Illustris and EAGLE simulations are completely independent from each other based on different codes and with different feedback recipes in galaxy evolution models \citep{Genel_2014,Vogelsberger_2014b,schaye2014eagle}. Dark matter halos are identified with the standard friends-of-friends (FOF) algorithm \citep{Davis_1985} and subhaloes within halos are detected with the Subfind algorithm \citep{Springel_2001, Dolag_2009}. Their physical properties are listed in the halo and subhalo catalogues, which can be downloaded from the Illustris and EAGLE webpages  \citep{Genel_2014,Vogelsberger_2014b,schaye2014eagle,Nelson_2018TNGrelease}. An overview about the physical and numerical parameters of the mentioned simulation runs are given in Table~\ref{tab:parameters}.
	
	\subsection{Illustris and IllustrisTNG simulations}
	The Illustris \citep{Nelson_2015Illustris} and IllustrisTNG \citep{Nelson_2018TNGrelease} simulations are performed with the moving-mesh code AREPO \citep{Springel_2010} and are based on a flat $\Lambda$CDM cosmology consistent with the Wilkinson Microwave Probe (WMAP)-9 \citep{Hinshaw_2013} and the Planck intermediate results \citep{ade2016planck}, respectively. We use the Illustris-1 and TNG100-1 simulations, of which each evolves $2 \times 1820^{3}$ particles or gas cells in a cube with a co-moving length of $75 \, h^{-1} \, \rm{cMpc}$  \citep{Vogelsberger_2014a,Nelson_2015Illustris,Nelson_2018TNGrelease}. In order to include also large-scale waves to get an accurate estimate on the peculiar velocity of subhaloes (see Section~\ref{sec:Velocity field of subhaloes}), we also employ the TNG300-1 simulation which has a co-moving box length of $205 \, h^{-1} \, \rm{cMpc}$. Both projects use different galaxy/physics models and compared to Illustris, IllustrisTNG includes magnetic fields (magnetohydrodynamical simulations) and has major modifications in the growth and feedback of supermassive black holes, galactic winds, and stellar evolution and gas chemical enrichment \citep{Marinacci_2018TNG, Naiman_2018TNG, Nelson_2018aTNG,Pillepich_2018TNG, Springel_2018TNG}. A detailed description about the IllustrisTNG galaxy models can be found in \citet{Pillepich_2018}.
	
	\subsection{EAGLE simulations}
	The EAGLE simulations \citep{schaye2014eagle} are performed with a modification of the GADGET-3 \citep{Springel_2005A} smoothed particle hydrodynamics code and with an implemented flat $\Lambda$CDM cosmology consistent with the initial Planck satellite data release \citep{Planck_2014}. Three different EAGLE simulation runs are used here as detailed in Table~\ref{tab:parameters}. EAGLE-1 is the largest simulation with a  co-moving length of $100 \, \rm{cMpc}$ per side and evolves $2 \times 1504^{3}$ particles across cosmic time \citep{schaye2014eagle}. EAGLE-2 and EAGLE-3 are high-resolution runs performed in a simulation box with a co-moving length of $25 \, \rm{cMpc}$ and $2 \times 752^{3}$ particles. The main difference between those two simulations is that in EAGLE-3 the stellar and active galactic nucleus feedback parameters are re-calibrated such that the galaxy stellar mass function fits the observations, i.e. EAGLE-2 predicts two times more galaxies $M_{*} > 10^{9} \, \rm{M_{\odot}}$ compared to observations of the local Universe \citep{schaye2014eagle,McAlpine_2016,Ploeckinger_2018}. 
	
	\begin{table*}
		\caption{Physical and numerical parameters of the Illustris-1, TNG100-1, and TNG300-1 simulations and different runs of the EAGLE project. $L$ is the co-moving length of the simulation box, $N$ is the total number of particles or cells, $H_{0}$ is the Hubble constant, $\Omega_{\mathrm{m}}$ is the total matter density, $\Omega_{\mathrm{b}}$ is the baryonic matter density, $\Omega_{\mathrm{\Lambda}}$ is the dark energy density at present time, and $m_{\mathrm{b}}$ and $m_{\mathrm{dm}}$ are the initial baryonic and dark matter particle masses, respectively. The gas cell resolution in Illsutris-1 is $48 \,\rm{pc}$ and in TNG100-1 (TNG300-1) the smallest gas cells have a radius of $14 \, \rm{pc}$ ($47 \, \rm{pc}$).}
		\label{tab:parameters}
		\begin{tabular}{lllllllll} \hline 
			Simulation & $L \, [\rm{cMpc}]$  & $N$ & $H_{0} \, [\rm{km \, s^{-1} \, Mpc^{-1}}]$ & $\Omega_{\mathrm{m}}$ & $\Omega_{\mathrm{b}}$ & $\Omega_{\mathrm{\Lambda}}$ & $m_{\mathrm{b}} \, [\rm{M_{\odot}}]$  & $m_{\mathrm{dm}} \, [\rm{M_{\odot}}]$  \\ \hline \hline 
			Illustris-1 & $106.5$ & $2 \times 1820^{3}$ & $70.4$  & $0.2726$ & $0.0456$ & $0.7274$ & $1.3 \times 10^{6}$ & $6.3 \times 10^{6}$ \\
			Illustris TNG100-1 & $110.7$ & $2 \times 1820^{3}$ & $67.74$  & $0.3089$ & $0.0486$ & $0.6911$ & $1.4 \times 10^{6}$ & $7.5 \times 10^{6}$ \\   
			Illustris TNG300-1 & $302.6$ & $2 \times 2500^{3}$ & $67.74$  & $0.3089$ & $0.0486$ & $0.6911$ & $1.1 \times 10^{7}$ & $5.9 \times 10^{7}$ \\ 
			EAGLE-1 Ref-L0100N1504 & $100$ & $2 \times 1504^{3}$ & $67.77$ & $0.307$ & $0.04825$ & $0.693$ & $1.81 \times 10^{6}$ & $9.70 \times 10^{6}$ \\
			EAGLE-2 Ref-L0025N0752& $25$ & $2 \times 752^{3}$ & $67.77$ & $0.307$ & $0.04825$ & $0.693$ & $2.26 \times 10^{5}$ & $1.21 \times 10^{6}$ \\
			EAGLE-3 Recal-L0025N0752 & $25$ & $2 \times 752^{3}$ & $67.77$ & $0.307$ & $0.04825$ & $0.693$ & $2.26 \times 10^{5}$ & $1.21 \times 10^{6}$ \\ \hline 
		\end{tabular}
	\end{table*}
	
	\subsection{Selection criteria and search-algorithm for DF2- and DF4-like dwarf galaxies} \label{sec:Methods_selection_criteria_algorithm}
	
	The here developed search-algorithm is sensitive to DF2- and DF4-like subhaloes based on their internal structures, masses, peculiar velocities, and motions relative to LG-like subhaloes scaled for different distances at redshift $z=0$. 
	
	Low-mass subhaloes can be substructures embedded in the galactic disc of a massive host galaxy \citep{Graus_2018, Ploeckinger_2018, Haslbauer_2019}. Therefore, the algorithm removes in the first step subhaloes which are within $10 \, \times$ the stellar half-mass radii of subhaloes being at least $10 \, \times$ more massive in stars. 
	NGC 1052 has an effective radius of $34$ arcsec \citep{Davies_1986} corresponding to a 3D deprojected half-light radius of $4.4 \, \rm{kpc}$ by assuming $D = 20 \, \rm{Mpc}$ and spherical geometry \citep{Wolf_2010}. The statistically expected 3D separation between the observed NGC 1052 and DF2 galaxies is about $98 \, \rm{kpc}$. Note that this analysis also includes non-central subhaloes.
	
	Secondly, the algorithm searches for substructure-corrected DF2-like subhaloes by following the description of \citet{vDokkum_2018a} according to which DF2 is located at a distance of $D = 20.0 \, \rm{Mpc}$. These subhaloes must have total-to-stellar mass ratios of $M_{\mathrm{total}}/M_{\mathrm*} < 2$ estimated by \citet{Danieli_2019}, stellar half-mass radii of either $R_{\mathrm{1/2}} \leq 2.7 \, \rm{kpc}$ or $R_{\mathrm{1/2}} \geq 2.7 \, \rm{kpc}$ depending on which of these two criteria gives, at the end, the lowest number of DF2-like objects (as is done in standard probability calculations), and $M_{*} = (4 \times 10^{7}-10^{9}) \, \rm{M_{\odot}}$. Since only the radial velocity of the observed DF2 dwarf galaxy (also for DF4) is known, one must take into account that this object can have an even higher $3$D velocity, with the statistically expected $3$D velocity being $\sqrt{3}$ times larger than its radial velocity. Each pre-selected subhalo is weighted by the fraction of all sky directions in which the peculiar velocities of selected subhaloes exceed the observed radial peculiar velocity of DF2. The sum over all weights, $w_{\mathrm{eff}}$, gives the actual effective number, $N_{\mathrm{eff}}$, of pre-selected DF2-like subhaloes in the simulation box 
	
	\begin{eqnarray}
	N_{\mathrm{eff}} \equiv \sum_{i}^{N} w_{\mathrm{eff},i} \equiv \sum_{i}^{N} \mathrm{max} \bigg( 0, ~1 - \frac{v_{\mathrm{DF2}}}{v_{\mathrm{pec},i}} \bigg) \, , 
	\label{eq:probability_peculiar_velocity}
	\end{eqnarray}
	where $N$ is the number of pre-selected DF2-like subhaloes, $v_{\mathrm{pec},i}$ is the magnitude of the peculiar velocity vector of the $i$-th subhalo wrt. the simulation box (i.e. CMB) and $v_{\mathrm{DF2}}$ is the observed distance-dependent radial peculiar velocity of DF2 given by 
	\begin{eqnarray}
	v_{\mathrm{DF2}} &=& v_{\mathrm{DF2,CMB}} - H_{0} D.
	\label{eq:peculiar_velocity_CMBframe_general}
	\end{eqnarray}
	Here, $D$ is the distance of DF2 from Earth and $v_{\mathrm{DF2, CMB}}$ is the peculiar velocity of DF2 wrt. the CMB reference frame being
	
	\begin{eqnarray}
	v_{\mathrm{DF2,CMB}} &=& \braket{v_{\mathrm{GCs}}} - v_{\mathrm{corr}} \\ 
	&=& \underbrace{\braket{v_{\mathrm{GCs}}}}_{\substack{\mathrm{DF2}}} -  \underbrace{(v_{\mathrm{helio}}  -  v_{\mathrm{CMB}})}_{\substack{\mathrm{NGC \, 1052}}} \, ,
	\label{eq:CMBframe}
	\end{eqnarray}
	where $\braket{v_{\mathrm{GCs}}} \approx 1803 \, \rm{km \, s^{-1}}$ is the mean velocity of $10$ GCs in DF2,\footnote{\citet{Emsellem_2018} found a slightly lower mean velocity of the diffuse stellar body of $1792.9_{+1.4}^{-1.8} \, \rm{km \, s^{-1}}$ compared to the mean velocity of $10$ GCs derived by \citet{vDokkum_2018a}.} $v_{\mathrm{helio}}$ is the heliocentric peculiar velocity, and $v_{\mathrm{CMB}}$ is the peculiar velocity wrt. the CMB reference frame. The CMB rest-frame correction, $v_{\mathrm{corr}}$, of DF2 is unknown and since DF2 has a similar sky position as NGC 1052, $v_{\mathrm{corr}}$ is calculated for NGC 1052 with $v_{\mathrm{helio}} = 1510 \pm 6 \, \rm{km \, s^{-1}}$ and $v_{\mathrm{CMB}} = 1293 \pm 16 \, \rm{km \, s^{-1}}$ taken from the NASA/IPAC Extragalactic Database\footnote{\url{https://ned.ipac.caltech.edu/}} \citep{Denicoloe_2005}. 
	
	In the third step, the motion of DF2 wrt. the LG is quantified. The algorithm searches for all pre-selected DF2-like subhalo LG-like objects, which are defined for simplicity as subhaloes with $M_{*} = 10^{10}-10^{12} \, \rm{M_{\odot}}$ and $M_{200} = 10^{12}-10^{13} \, \rm{M_{\odot}}$ within a spherical shell with $D \pm 1 \, \rm{Mpc}$ centred at the position of the pre-selected DF2-like subhalo. The line-of-sight velocity of the pre-selected DF2-like subhalo is
	
	\begin{eqnarray}
	\bm{v}_{\mathrm{DF2, los}} = \big( \bm{v}_{\mathrm{pec}} \cdot \widehat{\bm{r}} \big) \widehat{\bm{r}} \, ,
	\label{eq:los_peculiar_velocity}
	\end{eqnarray}
	where $\bm{v}_{\mathrm{pec}}$ is the velocity vector of the DF2-like subhalo and $\widehat{\bm{r}}$ is the normalized direction vector between an LG- and DF2-like galaxy. At $D = 20.0 \, \rm{Mpc}$, DF2 has a radial peculiar velocity of $186 \, \rm{km \, s^{-1}}$ towards $(l, b)_{\mathrm{DF2}} = (182.02^\circ, -57.93^\circ)$. DF4 has $-173 \, \rm{km \, s^{-1}}$ towards $(l, b)_{\mathrm{DF4}} = (181.27^\circ, -58.18^\circ)$ and is thus moving towards us. The relative peculiar velocity between LG- and DF2-like subhaloes is given by
	\begin{eqnarray}
	v_{\mathrm{rel}} = \left| \bm{v}_{\mathrm{DF2, los}} - \bm{v}_{\mathrm{LG}} \right| \, ,
	\label{eq:relative_peculiar_velocity}
	\end{eqnarray}
	where $\bm{v}_{\mathrm{LG}}$ is the peculiar velocity of LG-like subhaloes. The LG moves with a peculiar velocity of $\bm{v}_{\mathrm{LG}} = 627 \pm 22 \, \rm{km \, s^{-1}}$ wrt. the CMB frame towards $(l, b)_{\mathrm{LG}} = (276^\circ \pm 3^\circ, 30^\circ \pm 3^\circ)$ \citep{Kogut_1993}.
	The angle between the normalized vectors $\widehat{\bm{v}}_{\mathrm{DF2, los}}$ and $\widehat{\bm{v}}_{\mathrm{LG}}$ is calculated by
	
	\begin{eqnarray}
	\theta = \arccos \big( \widehat{\bm{v}}_{\mathrm{DF2, los}} \cdot \widehat{\bm{v}}_{\mathrm{LG}} \big) \, , 
	\label{eq:angle}
	\end{eqnarray}
	and is distance-independent. The observed angle between the peculiar velocity vector of the LG and the radial velocity vector of DF2 is $\theta = 117 \, ^{\circ}$.
	A pre-selected DF2-like subhalo is finally counted as a real DF2-like object if at least one simulated LG-DF2-like pair is found which fulfills both $v_{\mathrm{rel}} \geq v_{\mathrm{LG-DF2}}$ and $\theta \geq 117 \, ^{\circ}$, where $v_{\mathrm{LG-DF2}}$ is the observed relative velocity between the LG and DF2; otherwise the pre-selected DF2-like object is rejected and the effective number is reduced   
	
	\begin{eqnarray}
	N_{\mathrm{eff}} \rightarrow N_{\mathrm{eff}} - \sum_{j=1}^{N_{\mathrm{rej}}} w_{\mathrm{eff},j} \, , 
	\label{eq:new_effective_number}
	\end{eqnarray}
	where $N_{\mathrm{rej}}$ is the number of all rejected pre-selected DF2-like subhaloes.  
	Note that this approach is designed to minimize possible tension with $\Lambda$CDM because only one such LG-DF2-like pair is needed for the subhalo to count as a DF2-like object. The probability of DF2-like subhaloes, $p_{\mathrm{DF2,sim}}$, is then calculated by dividing the so-obtained effective number of DF2-like subhaloes by the number of substructure-corrected subhaloes within the stellar mass bin of $M_{*} = (4 \times 10^{7}-10^{9}) \, \rm{M_{\odot}}$. The probability of DF4-like subhaloes, $p_{\mathrm{DF4,sim}}$, is calculated in the same way. The structural probability of DF2-like subhaloes, $p_{\mathrm{DF2,structure}}$, is obtained by selecting substructure-corrected subhaloes without any constraints on the peculiar velocity and motion to LG-like subhaloes. In order to isolate the effect of the peculiar velocity constraints, we compare the effective number of analogs before and after applying them. Therefore, we define the peculiar velocity probability by
		\begin{eqnarray}
		 p_{\mathrm{DF2,vel}} \equiv \frac{p_{\mathrm{DF2,sim}}}{p_{\mathrm{DF2,structure}}} \, . 
		\label{eq:probability_sim_velocity}
		\end{eqnarray}
	
	The analysis is repeated for different distances down to $10.0 \, \rm{Mpc}$ with a resolution of $\Delta D = 0.5 \, \rm{Mpc}$ but only for the description of \citet{Martin_2018} in which $M_{\mathrm{total}}/M_{*}<8.1$ at $20.0 \, \rm{Mpc}$ is proposed in order to minimize tension with the $\Lambda$CDM model. The selection parameters $M_{*}$, $R_{1/2}$, $M_{\mathrm{total}}/M_{*}$, and velocities are adjusted for different distances, $D$, via the scaling relations stated in Appendix~\ref{appendix:scaling_relations}. In the Illustris-1, TNG100-1, and EAGLE-1 simulations $M_{*} = 10^{8} \, \rm{M_{\odot}}$ corresponds to $\ga 50$ stellar particles. The two high-resolution runs of the EAGLE simulations (EAGLE-2 and EAGLE-3) are also used to study subhaloes with $M_{*} = 10^{7} \, \rm{M_{\odot}}$ corresponding to $\approx 40$ stellar particles (see also Table~\ref{tab:parameters}). We choose a large stellar mass bin for selected DF2-like subhaloes in order to include subhaloes with much more stellar particles, i.e. at $10.0 \, \rm{Mpc}$ the upper end of the stellar mass bin is $2.5 \times 10^{8} \, \rm{M_{\odot}}$. We emphasize that DF2 has $M_{\mathrm{*}} = 2 \times 10^{8} \, \rm{M_{\odot}}$, which is sufficiently resolved for the here applied analysis. A very detailed description of the resolution issues is available in \citet{Haslbauer_2019}. It would be difficult to extend our analysis to $D < 10.0 \, \rm{Mpc}$ because the lower end of the stellar mass bin would be $M_{*} < 10^{7} \, \rm{M_{\odot}}$, making the resolution of subhaloes insufficient for the present analysis.
	
	DF4-like subhaloes with $M_{\mathrm{total}}/M_{*}<2$, a $R_{1/2}$ threshold of $ 2.0 \, \rm{kpc}$, $M_{*} = (4 \times 10^{7} - 10^{9}) \, \rm{M_{\odot}}$, and a mean radial velocity of the seven GCs of $\braket{v_{\mathrm{GCs}}} \approx 1446 \, \rm{km \, s^{-1}}$ are only analysed at $D = 20.0 \, \rm{Mpc}$.\footnote{The median radial velocity of the seven GCs of DF4 is $1445 \, \rm{km \, s^{-1}}$.} The selection criteria for DF2- and DF4-like galaxies scaled for different distances are summarized in Table~\ref{tab:selectioncriteria}.
	
	\begin{table*}
		\caption{Selection criteria for DF2- and DF4-like galaxies in the cosmological simulations by placing the observed DF2 galaxy at different distances, $D$, from Earth. The properties at $D=20 \, \rm{Mpc}$ are scaled to smaller distances using the scaling relations in Appendix~\ref{appendix:scaling_relations}.}
		\label{tab:selectioncriteria}
		\begin{tabular}{lllllllll} \hline 
			References & obj. & $D$ & $M_{\mathrm{total}}/M_{*}$  & $M_{*}$ & $R_{\mathrm{1/2}}$ & $v_{\mathrm{pec}}$ & $v_{\mathrm{rel}}$ & $\theta$ \\ 
			&  & $[\rm{Mpc}]$ & & $[\rm{M_{\odot}}]$ & $[\rm{kpc}]$ & $[\rm{km \, s^{-1}}]$ & $[\rm{km \, s^{-1}}]$ & $[^\circ]$\\ \hline \hline 
			\citet{vDokkum_2018a} & DF2 & $20$ & $< 2$ & $4.0 \times 10^{7}$ to $1.0 \times 10^{9}$ & $ 2.7$ & $\geq 186$ & $\geq 731$ & $\geq 117$ \\ \hline 
			\citet{Martin_2018} & DF2 & $20$ & $< 8.1$ & $4.0 \times 10^{7}$ to $1.0 \times 10^{9}$ & $ 2.7$ & $\geq 186$ & $\geq 731$ & $\geq 117$ \\
			$-$ & DF2 & $16$ & $< 10.1$ & $2.6 \times 10^{7}$ to $6.4 \times 10^{8}$ & $ 2.2$ & $\geq 466$ & $\geq 936$ & $\geq 117$ \\
			$-$ & DF2 & $13$ & $< 12.5$  & $1.7 \times 10^{7}$ to $4.2 \times 10^{8}$ & $ 1.8$ & $\geq 676$ & $\geq 1112$ & $\geq 117$ \\ 
			$-$ & DF2 & $10$ & $< 16.2$  & $1.0 \times 10^{7}$ to $2.5 \times 10^{8}$ & $ 1.4$ & $\geq 886$ & $\geq 1298$ & $\geq 117$ \\ \hline 
			\citet{vDokkum_2019} & DF4 & $20$ & $< 2$  & $4.0 \times 10^{7}$ to $1.0 \times 10^{9}$ & $ 2.0$ & $\geq 172$ & $\geq 723$ & $\geq 118$ \\ \hline 
		\end{tabular}
	\end{table*}

	\section{Results} \label{sec:Results}
	We estimate the probability of detecting DF2- and DF4-like galaxies in the most modern cosmological $\Lambda$CDM simulations at redshift $z=0$. The GC population in the observed DF2 dwarf galaxy is compared with that in the MW using the 2010 edition of the \citet{Harris_1996} catalogue.\footnote{\url{https://www.physics.mcmaster.ca/Fac_Harris/mwgc.dat}}
	
	\subsection{DF2- and DF4-like galaxies in cosmological simulations} \label{sec:Results_cosmological_simulations}

\begin{figure*}
		\centering
		\includegraphics[width=85mm]{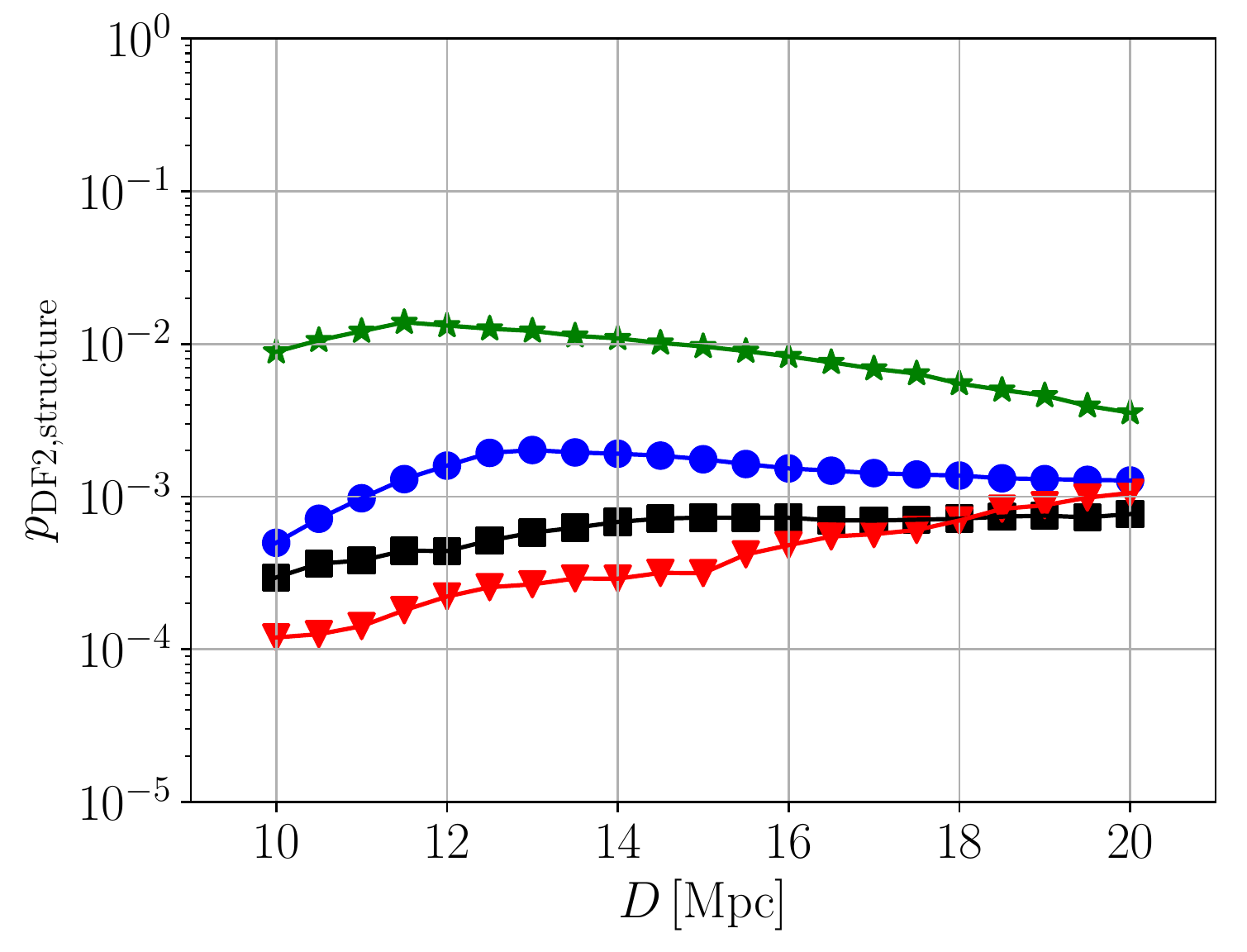}
		\includegraphics[width=85mm]{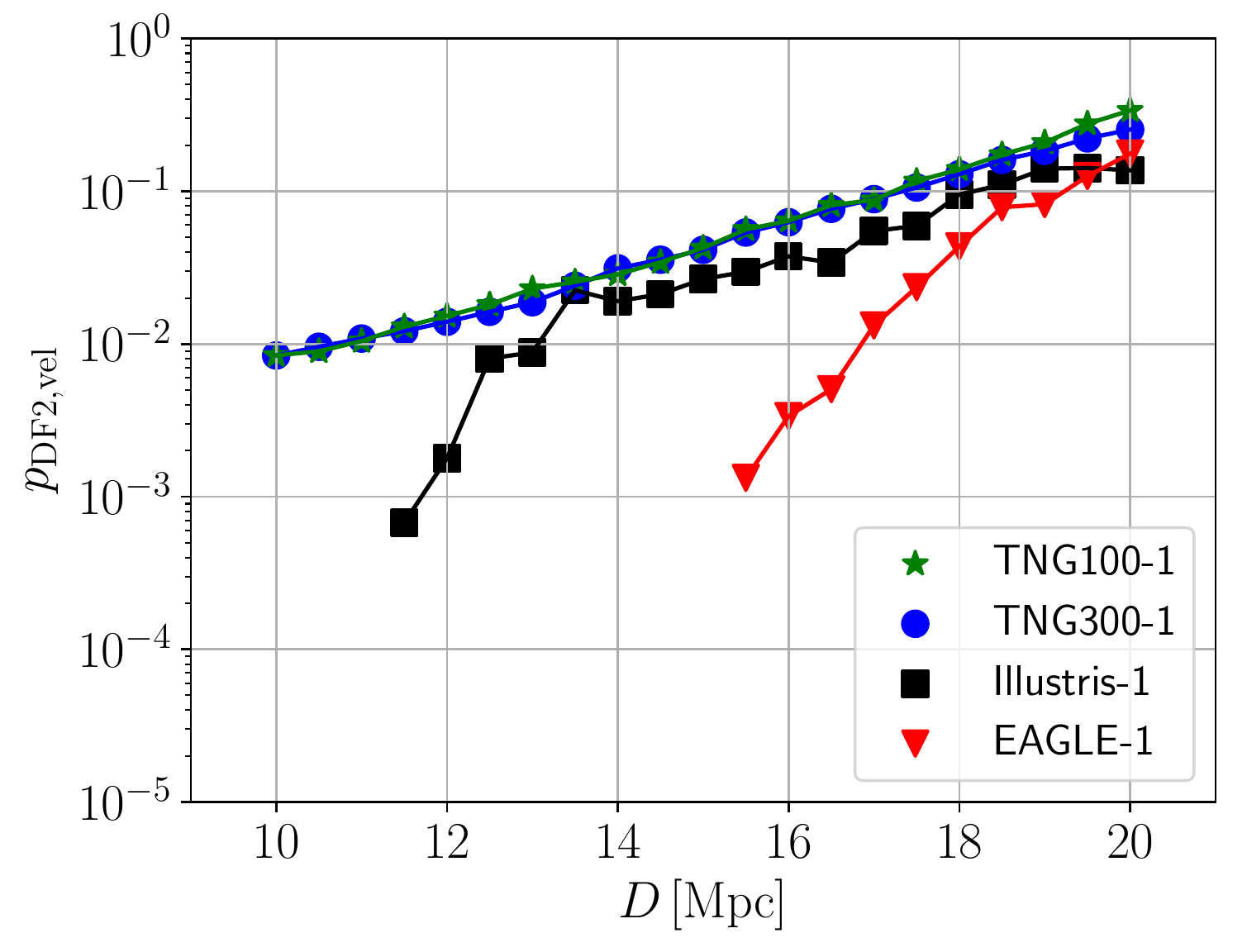}
		\caption{Left: Structural probability, $p_{\mathrm{DF2,structure}}$, of DF2-like subhaloes over distance, $D$, in the Illustris-1 (black squares), TNG100-1 (green stars), TNG300-1 (blue dots), and EAGLE-1 (red triangles) simulations. We note that these probabilities only include the structural properties of DF2 and do not include its peculiar velocity and GC population. Right: Peculiar velocity probability, $p_{\mathrm{DF2,vel}}$ (Equation~\ref{eq:probability_sim_velocity}), of DF2-like subhaloes over distance.}
		\label{fig:Pstructure}
	\end{figure*}

	The structural probability of the occurrence of DF2-like subhaloes in dependence of distance, selected based on its structural properties and assuming the physical description of \citet{Martin_2018}, is shown in the left-hand panel of Fig.~\ref{fig:Pstructure}. The constraints on $R_{1/2}$, $M_{*}$, and $M_{\mathrm{total}}/M_{*}$ cause a decrease of the frequency by about two orders of magnitude in the TNG100-1 simulation, in which the chance of finding such DF2-like galaxies becomes maximal at $D = 11.5_{-1.5}^{+4.0} \, \rm{Mpc}$. The role of the structural properties is discussed in more detail in Section \ref{sec:Structural properties of subhaloes}.
	
	The right-hand panel of Fig.~\ref{fig:Pstructure} reveals that, when the peculiar velocity is taken into account, the frequency of DF2-like subhaloes is reduced by another two orders of magnitude for $D \la 13 \, \rm{Mpc}$ and by about one order of magnitude for $D \ga 16 \, \rm{Mpc}$. The peculiar velocity constraints alone prefer the maximal allowed distance of $D=20.0 \, \rm{Mpc}$ with a $1 \sigma$ ($2 \sigma$) lower distance limit of $D=16.5 \, \rm{Mpc}$ ($D=12.0 \, \rm{Mpc}$). Had we allowed $D > 20 \, \rm{Mpc}$, $p_{\mathrm{vel}}$ would peak at an even larger distance $-$ the Hubble flow distance is $v_{\mathrm{DF2,CMB}}/H_{0} = 23 \, \rm{Mpc}$. At a distance of $13.0 \, \rm{Mpc}$ the observed DF2 galaxy would have a high peculiar velocity wrt. the CMB of $v_{\mathrm{pec}} \approx 676 \, \rm{km \, s^{-1}}$, which differs from that of the LG by $v_{\mathrm{rel}} \approx 1112 \, \rm{km \, s^{-1}}$ with an angle of $\theta = 117 \, ^{\circ}$. About $13 \, \%$ of pre-selected DF2-LG-like systems fulfill the latter two constraints and according to Table~\ref{tab:LG_statistics} this fraction shows only a week dependence with distance for the TNG300-1 simulation. Thus, the decrease of the frequency of DF2-like galaxies for smaller distance is generated by the peculiar velocity of DF2 wrt. the CMB reference frame. 
	
	The distance-dependent peculiar velocity probability is almost exactly the same for subhaloes in the TNG100-1 and TNG300-1, which is spatially three times larger than the former simulation run and includes therefore much more large-scale waves. This consistency between these two simulation runs could be either a pure coincidence or it means that the peculiar velocity field of DF2-like subhaloes has already converged in the TNG100-1 simulation. The convergence of the velocity fields of TNG100-1 and TNG300-1 are addressed in Section \ref{sec:Velocity field of subhaloes}. Since TNG100-1 has a much better resolution than TNG300-1 (see Table \ref{tab:parameters}) and the role of the peculiar velocity is the same for both simulations, we use the former one for our main results.
	
\begin{figure}
		\centering
		\includegraphics[width=85mm]{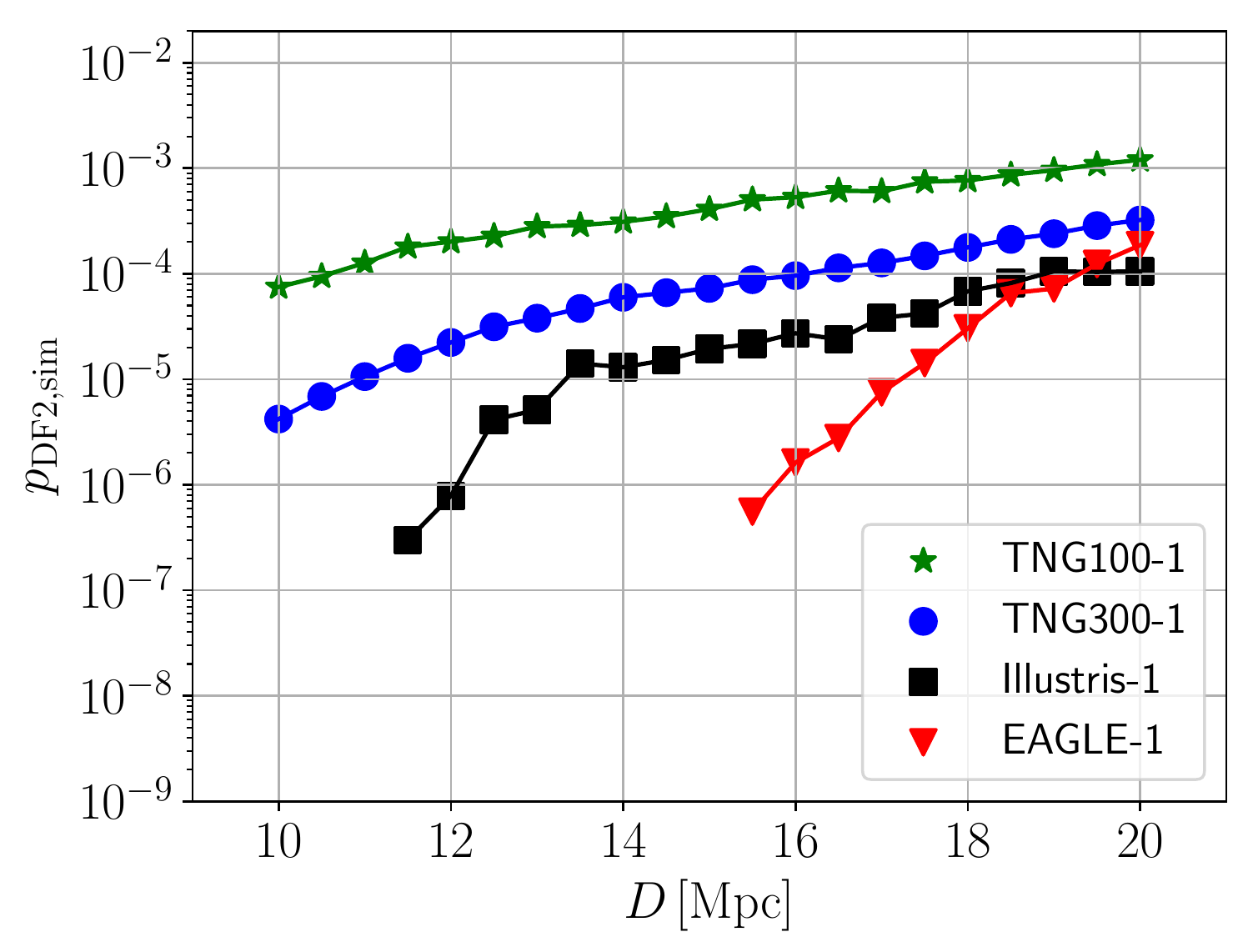}
		\caption{Cosmological detection probability, $p_{\mathrm{DF2,sim}}$, in dependence of its distance, $D$, from Earth for the Illustris-1 (black squares), TNG100-1 (green stars), TNG300-1 (blue dots), and EAGLE-1 (red triangles) simulations. There are no DF2-like subhaloes in the EAGLE-2 and EAGLE-3 simulations.}
		\label{fig:Psim_all}
	\end{figure}

	The probability of DF2-like galaxies occurring in a $\Lambda$CDM cosmology evaluated by combining the structural and peculiar velocity constraints without considering its GC population is finally given in Fig.~\ref{fig:Psim_all}. The probability increases for a larger distance and becomes maximal for our allowed distance range in TNG100-1 with $1.2 \times 10^{-3}$ at a distance of $D = 20.0 \, \rm{Mpc}$ with a $1\sigma$ ($2 \sigma$) lower distance limit of $16.0 \, \rm{Mpc}$ ($11.5 \, \rm{Mpc}$). The effective numbers, co-moving number densities, and probabilities of detecting DF2-like galaxies with $M/L$ ratios consistent with \citet{Martin_2018} for different distances and simulation runs are listed in Table~\ref{tab:results_all_distances}. 
	
	\begin{table}
		\caption{Fraction of pre-selected DF2 analogues for which an LG-like vantage point exists with $v_{\mathrm{rel}} \geq v_{\mathrm{LG-DF2}}$ (Equation~\ref{eq:relative_peculiar_velocity}) and $\theta \geq 117 \, ^{\circ}$ (Equation~\ref{eq:angle}) for different adopted distances in the TNG300-1 simulation ($D = 10, 13, 16, 20 \, \rm{Mpc}$ and $v_{\mathrm{LG-DF2}} = 1298, 1112, 936, 731 \, \rm{km \, s^{-1}}$, respectively).}
		\label{tab:LG_statistics}
		\begin{tabular}{llllllllllll} \hline 
			$D \, [\rm{Mpc}]$ & $v_{\mathrm{rel}} \geq v_{\mathrm{LG-DF2}}$ & $\theta \geq 117 \, ^{\circ}$ & $\theta \geq 117 \, ^{\circ} \land v_{\mathrm{rel}} \geq v_{\mathrm{LG-DF2}}$  \\ \hline \hline 
			$20$ & $0.22$ & $0.21$ & $0.094$ \\ 
			$16$ & $0.20$ & $0.22$ & $0.11$  \\ 
			$13$ & $0.20$ & $0.26$ & $0.13$  \\ 
			$10$ & $0.22$ & $0.29$ & $0.14$  \\ \hline 
		\end{tabular}
	\end{table}

	\begin{table*}
		\caption{Effective number, $N_{\mathrm{DF2,eff}}$ (Equation~\ref{eq:probability_peculiar_velocity} and Equation~\ref{eq:new_effective_number}), effective co-moving number density, $n_{\mathrm{DF2,eff}}$, and fraction of DF2-like galaxies, $p_{\mathrm{DF2,sim}}$, in different simulation runs for different distances, $D$. The numbers in brackets give the unweighted number of DF2-like subhaloes, $N_{\mathrm{DF2}}$. Here we assume $M_{\mathrm{total}}/M_{*}$ ratios consistent with \citet{Martin_2018}. The selection criteria for DF2- and DF4-like galaxies are listed in Table~\ref{tab:selectioncriteria}.}
		\label{tab:results_all_distances}
		\tiny
		\begin{tabular}{lllllllllllll} \hline 
			Distances & \multicolumn{3}{l}{$20 \, \rm{Mpc}$ \citep{Martin_2018}} & $16 \, \rm{Mpc}$  & & & $13 \, \rm{Mpc}$ & & & $10 \, \rm{Mpc}$   \\ \hline  
			Simulation & $N_{\mathrm{DF2,eff}}$ ($N$) & $n_{\mathrm{DF2,eff}}$ & $p_{\mathrm{DF2,sim}}$ & $N_{\mathrm{DF2,eff}}$ ($N$) & $n_{\mathrm{DF2,eff}}$ & $p_{\mathrm{DF2,sim}}$ & $N_{\mathrm{DF2,eff}}$  ($N$) & $n_{\mathrm{DF2,eff}}$ & $p_{\mathrm{DF2,sim}}$ & $N_{\mathrm{DF2,eff}}$ ($N$) & $n_{\mathrm{DF2,eff}}$ & $p_{\mathrm{DF2,sim}}$ \\
			& & $[\rm{cMpc^{-3}}]$ & & & $[\rm{cMpc^{-3}}]$ & & & $[\rm{cMpc^{-3}}]$ & & & $[\rm{cMpc^{-3}}]$ &  \\ \hline \hline  
			Illustris-1 & $7.5$ $(12)$ & $6.2 \times 10^{-6}$ & $1.1 \times 10^{-4}$ & $2.2$ $(5)$ & $1.8 \times 10^{-6}$ & $2.7 \times 10^{-5}$ & $0.45$ $(2)$ & $3.7 \times 10^{-7}$ & $5.1 \times 10^{-6}$ & $0$ $(0)$ & $0$ & $0$\\ 
			TNG100-1    & $52.0$ $(79)$ & $3.8 \times 10^{-5}$ & $1.2 \times 10^{-3}$ & $26.2$ $(69)$ & $1.9 \times 10^{-5}$ & $5.3 \times 10^{-4}$ & $15.7$ $(51)$ & $1.2 \times 10^{-5}$ & $2.8 \times 10^{-4}$ & $5.1$ $(20)$ & $3.7 \times 10^{-6}$ & $7.5 \times 10^{-5}$\\ 
			TNG300-1  & $180.3$ $(275)$ & $6.5 \times 10^{-6}$ & $3.2 \times 10^{-4}$ & $65.6$ $(163)$ & $2.4 \times 10^{-6}$ & $9.6 \times 10^{-5}$ & $31.9$ $(102)$ & $1.2 \times 10^{-6}$ & $3.8 \times 10^{-5}$ & $4.5$ $(15)$ & $1.6 \times 10^{-7}$ & $4.2 \times 10^{-6}$\\ 
			EAGLE-1     & $9.2$ $(16)$ & $9.2 \times 10^{-6}$ & $1.9 \times 10^{-4}$ & $0.10$ $(1)$ & $1.0 \times 10^{-7}$ & $1.6 \times 10^{-6}$ & $0$  $(0)$& $0$ & $0$ & $0$ $(0)$ & $0$ & $0$\\ 
			EAGLE-2     & $0$ $(0)$ & $0$ & $0$ & $0$ $(0)$ & $0$ & $0$ & $0$  $(0)$ & $0$ & $0$ & $0$  $(0)$ & $0$ & $0$\\  
			EAGLE-3     & $0$ $(0)$ & $0$ & $0$ & $0$ $(0)$ & $0$ & $0$ & $0$  $(0)$ & $0$ & $0$ & $0$  $(0)$ & $0$  & $0$\\ 
			\hline 
		\end{tabular}
	\end{table*}

	\begin{table*}
		\caption{Same as Table~\ref{tab:results_all_distances} but for DF2- and DF4-like galaxies if located at a distance of $20 \, \rm{Mpc}$. We assume for DF2 different $M_{\mathrm{total}}/M_{*}$ ratios (i.e. $M_{\mathrm{total}}/M_{*} < 8.1$ \citealt{Martin_2018} and $M_{\mathrm{total}}/M_{*} < 2$ \citealt{vDokkum_2018a}; see also Table~\ref{tab:selectioncriteria}).}
		\label{tab:results_20Mpc}
		\tiny
		\begin{tabular}{lllllllllllllll} \hline 
			References & \multicolumn{3}{l}{$20 \, \rm{Mpc}$ \citep{Martin_2018}} & \multicolumn{3}{l}{$20 \, \rm{Mpc}$ \citep{vDokkum_2018a}} & \multicolumn{3}{l}{$20 \, \rm{Mpc}$ \citep{vDokkum_2019}} \\ \hline  
			Simulation & $N_{\mathrm{DF2,eff}}$ ($N$) & $n_{\mathrm{DF2,eff}} \, [\rm{cMpc^{-3}}]$ & $p_{\mathrm{DF2,sim}}$& $N_{\mathrm{DF2,eff}}$ ($N$) & $n_{\mathrm{DF2,eff}} \, [\rm{cMpc^{-3}}]$ & $p_{\mathrm{DF2,sim}}$ & $N_{\mathrm{DF4,eff}}$ ($N$) & $n_{\mathrm{DF4,eff}} \, [\rm{cMpc^{-3}}]$ & $p_{\mathrm{DF4,sim}}$  \\ \hline \hline  
			Illustris-1 & $7.5$ $(12)$ & $6.2 \times 10^{-6}$ & $1.1 \times 10^{-4}$ & $2.9$  $(5)$& $2.4 \times 10^{-6}$ & $4.1 \times 10^{-5}$ & $3.1$ $(5)$& $2.6 \times 10^{-6}$ & $4.3 \times 10^{-5}$  \\  
			TNG100-1 & $52.0$ $(79)$ & $3.8 \times 10^{-5}$ & $1.2 \times 10^{-3}$ & $3.9$ $(6)$ & $2.9 \times 10^{-6}$& $9.0 \times 10^{-5}$ & $4.7$ $(7)$& $3.5 \times 10^{-6}$ & $1.1 \times 10^{-4}$  \\  
			TNG300-1  & $180.3$ $(275)$ & $6.5 \times 10^{-6}$ & $3.2 \times 10^{-4}$ & $13.8$ $(20)$ & $5.0 \times 10^{-7}$ & $2.5 \times 10^{-5}$ & $80.2$ $(117)$ & $2.9 \times 10^{-6}$ & $1.4 \times 10^{-4}$  \\ 
			EAGLE-1  & $9.2$ $(16)$ & $9.2 \times 10^{-6}$ & $1.9 \times 10^{-4}$ & $0$ $(0)$ & $0$ & $0$ & $0$ $(0)$ & $0$ & $0$ \\  
			EAGLE-2 & $0$ $(0)$ & $0$ & $0$ & $0$ $(0)$ & $0$ & $0$ & $0$ $(0)$ & $0$ & $0$  \\ 
			EAGLE-3 & $0$ $(0)$ & $0$ & $0$ & $0$ $(0)$ & $0$ & $0$ & $0$ $(0)$ & $0$ & $0$  \\ 
			\hline 
		\end{tabular}
	\end{table*}

	By adopting the physical properties of both galaxies according to \citet{vDokkum_2018a, vDokkum_2019} the maximal probability for finding each DF2- and DF4-like object is $9.0 \times 10^{-5}$ (TNG100-1) and $1.4 \times 10^{-4}$ (TNG300-1), respectively. That is, to find both at the same time has a probability of $1.3 \times 10^{-8}$. Therefore, in order to minimize the tensions with $\Lambda$CDM theory our main analysis relies on the description of \citet{Martin_2018} according to which the $M_{\mathrm{total}}/M_{*}$ ratio is about four times larger than proposed by \citet{vDokkum_2018a}. The effective numbers, co-moving number densities, and probabilities of DF2- and DF4-like galaxies assuming $D = 20 \, \rm{Mpc}$ are summarized in Table~\ref{tab:results_20Mpc}.
	
	Thus, applying the above-stated selection criteria (see Table~\ref{tab:selectioncriteria}) on the subhaloes of the Illustris-1, TNG100-1, TNG300-1, and EAGLE simulations shows that both DF2- and DF4-like galaxies are rare in $\Lambda$CDM simulations. Neither DF2 nor DF4 analogues can be found in the EAGLE-2 and EAGLE-3 simulations. In the following section we quantify the distance-dependent likelihood of the GC population and combine it with the probability obtained from the cosmological simulations.
	  
	\subsection{Luminosities of globular clusters in DF2} \label{sec:Results_Luminosities}
	The observed DF2 dwarf galaxy has unusually bright and large GCs if located at a distance of around $20 \, \rm{Mpc}$ from Earth. This is notable, because up to now it is understood that all galaxies have a canonical peak of the GC luminosity distribution at $M_{\mathrm{V}} = -7.5$ \citep{Rejkuba_2012}. In contrast to that, \citet{vDokkum_2018d} estimate that DF2 has in total $15$ GCs with absolute magnitudes $M_{\mathrm{V,606}} < -6.5$ and $11$ of them are spectroscopically confirmed and have $M_{\mathrm{V,606}} \leq -8.6$, therewith being much brighter than the expected canonical peak from other observed galaxies. This exceptional GC population poses the question if DF2 might be located at a much smaller distance from Earth, which will be addressed by the following statistical tests. We scale the luminosities of $15$ GCs down to smaller distances and compare them with those of the MW\footnote{The Harris catalogue \citep{Harris_1996} lists the absolute visual magnitudes of GCs in the MW which can be converted to the $V_{606}$ band via $M_{\mathrm{V,606}} = M_{\mathrm{V}} - 0.05$.} \citep{Harris_1996} within the same luminosity ranges starting with $M_{\mathrm{V,606}} \leq -6.5$ at $D = 20 \, \rm{Mpc}$. The right and bottom panels of fig.~3 in \citet{vDokkum_2018d} show the observed interloper-corrected luminosity function of the compact objects (GCs) in DF2 according to which almost no additional GCs can be found in the $-5 < M_{V,606} < - 6.5$ range. This allows us to scale the magnitudes (and sizes) of the $15$ GCs up to $D = 10 \, \rm{Mpc}$.
	
	Since the magnitudes of the four GCs within $-8.6< M_{\mathrm{V,606}} < -6.5$ are not reported in \citet{vDokkum_2018d}, we assume that all of them have $M_{\mathrm{V,606}} = -7.5$ to minimize the tension with the GC population of other observed galaxies. 
	The $P$-values of the following statistical tests on the GC luminosity distributions scaled for different distances are listed in Table~\ref{tab:results_statistics_luminosities}.  
	
	\subsubsection{Binomial distribution} \label{sec:Results_Luminosities_Binomial}
	
	Given that observed quantities, $X$, are binomially distributed, which we symbolically express here by $X \sim B(n, p)$, the corresponding cumulative distribution function has the form
	
	\begin{eqnarray}
	P(\geq  k;p,n) \equiv P(X \geq k \vert X \sim B(n, p)) = \sum_{i=k}^{  n } \binom{n}{i} p^{i} (1-p)^{n-i} \, ,
	\label{eq:binomial_general}
	\end{eqnarray}
	where 
	\begin{eqnarray}
	\binom{n}{k} = \frac{n!}{k!(n-k)!}
	\label{eq:binomial_coefficient}
	\end{eqnarray}
	is the binomial coefficient, $n$ is the number of trials ($n \in  \mathbb{N}$), $k$ is the number of successes ($k \in [0,n]$), and $p^{k}$ is the probability of having $k$ successes ($p \in [0,1]$).
	
	Assuming that the magnitudes of a GC population are binomially distributed, we consider two different approaches to quantify the differences between the GC populations of the MW and DF2. The first method (hereafter method $1$) is sensitive to the medians of the GC magnitudes. We set $p = 0.5$, $n = 15$ (number of GCs in DF2), and $k$ is the number of DF2 GCs smaller or equal than the median of the magnitude of the GCs in the MW with $M_{V,606} \leq -6.5$ if $D = 20 \, \rm{Mpc}$.  
	
	The second method (method $2$) focuses more on the brightest GCs of the distribution. Here, $p$ is the fraction of MW GCs with $M_{V,606} \leq -8.6$ and $M_{V,606} \leq -6.5$, $n=15$, and $k$ is the number of GCs in DF2 with $M_{V,606} \leq -8.6$ if located at $D = 20 \, \rm{Mpc}$. Note that these magnitude thresholds are adopted for different distances with the scaling relation described in Appendix~\ref{appendix:scaling_relations}. 
	
	\subsubsection{Comparison of the means} \label{sec:Results_Luminosities_means}
	The null hypothesis states that the means of two samples have to be the same. Since the Harris catalogue and table~1 in \citet{vDokkum_2018d} do not list the uncertainties of the GC magnitudes, we have to assume that the error on the mean of magnitudes of the MW GCs is zero and the uncertainty of the DF2 GCs is given by the intrinsic scatter of the MW GC population, i.e.  $\sigma_{\mathrm{MW}}/\sqrt{N}$, where $N$ is the number of GCs in DF2 and $\sigma_{\mathrm{MW}}$ is the standard deviation of the magnitudes of the MW GCs. 
	
	In the case of two normal distributions, the standard score of the comparison of the means is then given by
	\begin{eqnarray}
	Z =  \left| \mu_{\mathrm{DF2}}-\mu_{\mathrm{MW}} \right| \div \frac{\sigma_{\mathrm{MW}}}{\sqrt{N}} \, ,
	\label{eq:comparison_means_luminosity}
	\end{eqnarray}
	where $\mu_{\mathrm{MW}}$ and $\mu_{\mathrm{DF2}}$ are the means of the magnitudes of the MW and DF2 GCs. The two-tailed $P$-value is calculated by using a Gaussian distribution. 
	
	\subsubsection{Mann--Whitney U test} \label{sec:Results_Luminosities_Utest}
	The Mann--Whitney U test \citep{MannWhitney_1947}, also known as the Wilcoxon 2-sample test \citep[e.g.][]{Kroupa_1995}, is a non-parametric test, which compares the medians of two populations whereby the null hypothesis states that their medians are equal. The U test statistic for a two-tailed hypothesis is calculated by 
	
	\begin{eqnarray}
	U &=& \mathrm{max} (u_{1}, u_{2}) \, , 
	\label{eq:Utest_1}
	\end{eqnarray}
	with
	\begin{eqnarray}
	u_{1} &=& n_{1} n_{2} + \frac{n_{1} (n_{1} + 1) }{2} - \sum_{i = 1}^{n_{1}} r_{1,i} \, , \\
	u_{2} &=& n_{1} n_{2} - u_{1} \, , 
	\label{eq:Utest_2}
	\end{eqnarray}
	where $n_{1}$ and $n_{2}$ are the sample sizes and $r_{1}$ are the ranks of sample $1$. If the samples are large enough we can assume that U follows a normal distribution, in which the standard score is obtained by
	\begin{eqnarray}
	\mu &=& \frac{n_{1} n_{2}}{2} +\underbrace{0.5}_{\substack{C}} \, ,\\ 
	\sigma_{\mathrm{corr}} &=& \sqrt{T_{\mathrm{corr}} n_{1} n_{2} \bigg( \frac{n + 1}{12} \bigg) }  \, , \\ 
	T_{\mathrm{corr}} &=& 1 - \sum_{i=0}^{k} \bigg( \frac{t_{i}^{3} - t_{i} }{n^{3} - n} \bigg)  \, , \\ 
	Z &=& \frac{U-\mu}{\sigma_{\mathrm{corr}}}  \, , 
	\label{eq:Utest_3}
	\end{eqnarray}
	where $n = n_{1} + n_{2}$, $C$ is a continuity correction, $T_{\mathrm{corr}}$ is the tie correction function, $t_{i}$ is the number of subjects sharing rank $i$, and $k$ is the number of tied ranks. 
	
	\subsubsection{Kolmogorov--Smirnov test} \label{sec:Results_Luminosities_KStest} 
\begin{figure}
		\centering
		\includegraphics[width=\linewidth]{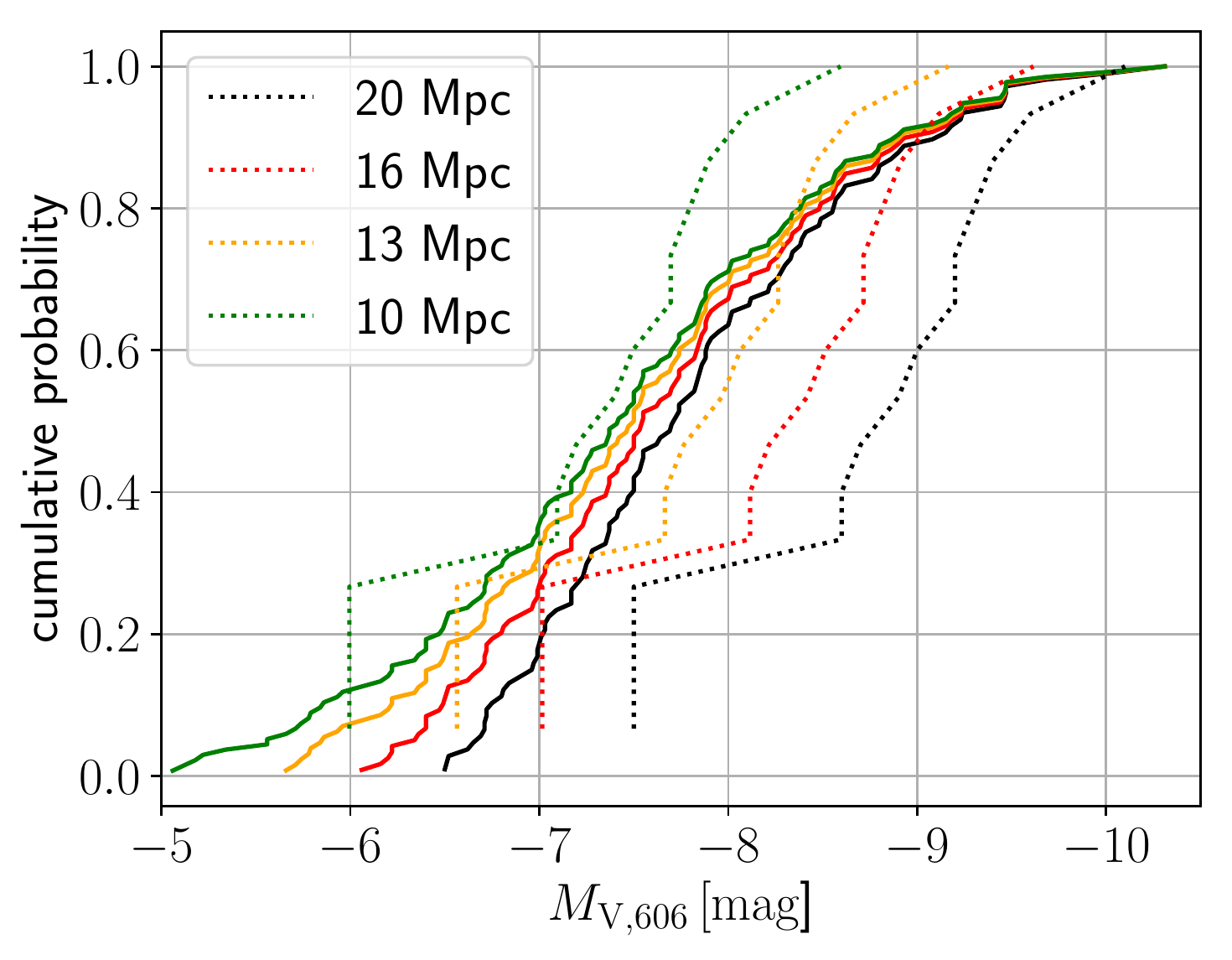}
		\caption{Cumulative probability distributions of the GC magnitude distribution in the MW (solid lines) and DF2 (dashed lines). Different line colours refer to different distances of the observed DF2 galaxy from Earth (from left to right: $D = 10 \, \rm{Mpc}$, $13 \, \rm{Mpc}$, $16 \, \rm{Mpc}$, and $20 \, \rm{Mpc}$). The maximal differences between those two distributions are $0.21$ ($-8.1 \, \rm{mag}$), $0.30$ ($-7.7 \, \rm{mag}$), $0.43$ ($-8.1 \, \rm{mag}$), and $0.55$ ($-8.6 \, \rm{mag}$), respectively. Note that the four DF2 GCs with $M_{\mathrm{V, 606}} > -8.6$ are assumed to have $M_{\mathrm{V,606}}=-7.5$.}
		\label{fig:KS_test_mag}
	\end{figure}

	The Kolmogorov--Smirnov (KS) test is also a non-parametric test comparing the cumulative probability distribution of two independent samples. The null hypothesis here is that the GC population of DF2 is drawn from that of the MW. The $P$-value is calculated by
	
	\begin{eqnarray}
	m &=& \frac{n_{1} n_{2}}{n_{1} + n_{2}}  \, , \\
	\lambda &=& \bigg(\sqrt{m} + 0.12 + \frac{0.11}{\sqrt{m}} \bigg) D_{\mathrm{max}} \, , \\
	P &=& \sum_{i=1}^{\infty} (-1)^{i+1} \exp\big({-2 \lambda^{2} i^{2}}\big) \, ,
	\label{eq:KStest}
	\end{eqnarray}
	where $n_{1}$ and $n_{2}$ are the numbers of the GCs in the MW and DF2, respectively, and $D_{\mathrm{max}}$ is the maximal difference in their cumulative probability distributions. The cumulative probability distributions of the GC magnitude distributions in the MW and DF2 scaled for different distances and their corresponding $D_{\mathrm{max}}$ values are presented in Fig.~\ref{fig:KS_test_mag}. Compared to the other above statistical test, the KS test is more sensitive to the whole GC population and is commonly used when the distributions of two samples have to be compared. Therefore, when calculating the combined probability of DF2 analogues in standard cosmology based on its occurrence in cosmological simulations and its GC population, the $P$-values of the KS test are used (see Section~\ref{sec:combined_probability}). The other tests (Sections~\ref{sec:Results_Luminosities_Binomial}--\ref{sec:Results_Luminosities_Utest}) are applied as consistency checks.
	
	\begin{table*}
		\caption{$P$-values of the binomial cumulative functions (see methods 1 and 2 in 
			Section~\ref{sec:Results_Luminosities_Binomial}), comparison of the means (see Section~\ref{sec:Results_Luminosities_means}), Mann--Whitney U test (see Section~\ref{sec:Results_Luminosities_Utest}), and KS test (see Section~\ref{sec:Results_Luminosities_KStest}) applied on the magnitude distribution of the GC population in the MW and DF2.}
		\label{tab:results_statistics_luminosities}
		\begin{tabular}{llllllll} \hline 
			$D \, [\rm{Mpc}]$ & $M_{V,606}$ & $P$-value (binomial 1) & $P$-value (binomial 2) & $P$-value (mean comp.) & $P$-value (U-test)  & $P$-value (KS test)  \\ \hline \hline 
			$20$ & $\leq -6.5$ & $5.9 \times 10^{-2}$ & $6.3 \times 10^{-6}$ & $7.3 \times 10^{-5}$ & $8.3 \times 10^{-4}$ & $4.0 \times 10^{-4}$ \\ 
			$16$ & $\leq -6.0$ & $5.9 \times 10^{-2}$ & $7.3 \times 10^{-4}$ & $2.4 \times 10^{-2}$ & $2.6 \times 10^{-2}$ & $9.4 \times 10^{-3}$ \\
			$13$ & $\leq -5.6$ & $5.9 \times 10^{-2}$ & $2.0 \times 10^{-2}$ & $3.9 \times 10^{-1}$ & $3.0 \times 10^{-1}$ & $1.6 \times 10^{-1}$ \\
			$10$ & $\leq -5.0$ & $7.0 \times 10^{-1}$ & $2.3 \times 10^{-1}$ & $4.3 \times 10^{-1}$ & $5.4 \times 10^{-1}$ & $5.6 \times 10^{-1}$ \\ \hline 
		\end{tabular}
	\end{table*}
	
	\subsection{Half-light radii of globular clusters in DF2} \label{sec:Results_Radii}
	Here we apply the same statistical tests of Section~\ref{sec:Results_Luminosities} on the GC sizes but only to the $11$ GCs with $M_{V, 606} \leq -8.6$ since for those the half-light radii and also their uncertainties are reported in \citet{vDokkum_2018d}. Thus, at $D = 20 \, \rm{Mpc}$ a luminosity range of $M_{V,606} \leq -8.6$ is chosen, which scales to $M_{V,606} \leq -7.1$ at $10 \, \rm{Mpc}$. 
	
\begin{figure}
		\centering
		\includegraphics[width=\linewidth]{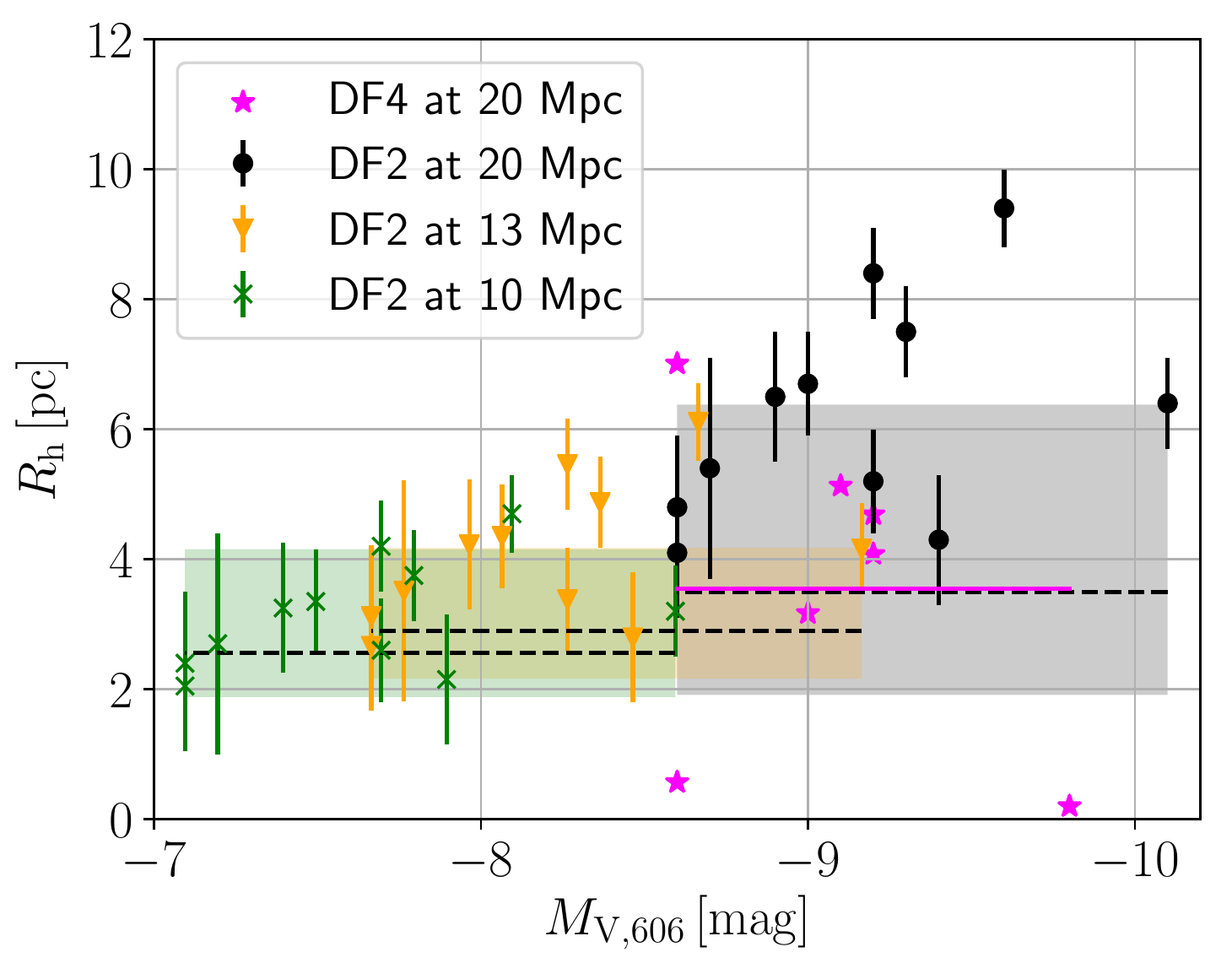}
		\caption{Circularized half-light radius, $R_{\mathrm{h}}$, in dependence of the absolute magnitude, $M_{\mathrm{V,606}}$, of GCs in DF2 and DF4 (magenta) with $M_{\mathrm{V,606}} \leq -8.6$ at $20 \, \rm{Mpc}$ and scaled for different distances. The dashed horizontal lines mark the median of the half-light radii of GCs in the MW in the respective luminosity range and the shaded area marks the $16^{\mathrm{th}}$ and $84^{\mathrm{th}}$ percentile. The magenta solid line highlights the median of half-light radii of the seven GCs in DF4 \citep{Danieli_2019}. The data points at $20 \, \rm{Mpc}$ and errorbars  are taken from \citet{vDokkum_2018d} (see table~1 and fig.~4).}
		\label{fig:GCs}
	\end{figure}

	Fig.~\ref{fig:GCs} presents the absolute magnitudes and sizes of GCs with $M_{V, 606} \leq -8.6$ in DF2 and DF4 if located at $D=20 \, \rm{Mpc}$. The properties of the DF2 GCs are scaled to smaller distances and compared with MW GCs within the same luminosity range. Table~\ref{tab:results_statistics_radii} summarizes the $P$-values of the statistical tests on the GC half-light radius distributions. 
	
	\subsubsection{Binomial distribution} \label{sec:Results_Radii_Binomial}
	
	Same as method~$1$ in Section~\ref{sec:Results_Luminosities_Binomial} with $p = 0.5$ and $n = 11$ by assuming that the GC sizes are binomially distributed. 
	
	\subsubsection{Comparison of the means} \label{sec:Results_Radii_means}
	
	Table~$1$ in \citet{vDokkum_2018d} lists the uncertainties of the DF2 GC sizes. Thus, in contrast to Section~\ref{sec:Results_Luminosities_means} we have to use the inverse-variance weighted average
	\begin{eqnarray}
	\sigma^2 &=& 1 \div \sum_{i} \frac{1}{(\sigma_{i}^{2}+\sigma_{\mathrm{MW}}^{2})} \, , \\
	\bar{\mu}_{\mathrm{DF2}} &=& \sigma^2 \sum_{i} \frac{R_{\mathrm{h},i}}{(\sigma_{i}^{2} + \sigma_{\mathrm{MW}}^{2})} \, ,
	\label{eq:comparison_means_radii}
	\end{eqnarray}
	where $R_{\mathrm{h},i}$ and $\sigma_{i}$ are the half-light radius and corresponding uncertainty of the $i$-th DF2 GC, and $\sigma_{\mathrm{MW}}$ is the standard deviation of the MW GC radii within the same luminosity range. Again, the standard score and the two-tailed $P$-value are evaluated.
	
	\subsubsection{Mann--Whitney U test} \label{sec:Results_Radii_Utest}
	As in Section~\ref{sec:Results_Luminosities_Utest} but applied on the GC half-light radii. 
	
	\subsubsection{Kolmogorov--Smirnov test} \label{sec:Results_Radii_KStest}
	
\begin{figure}
		\centering
		\includegraphics[width=\linewidth]{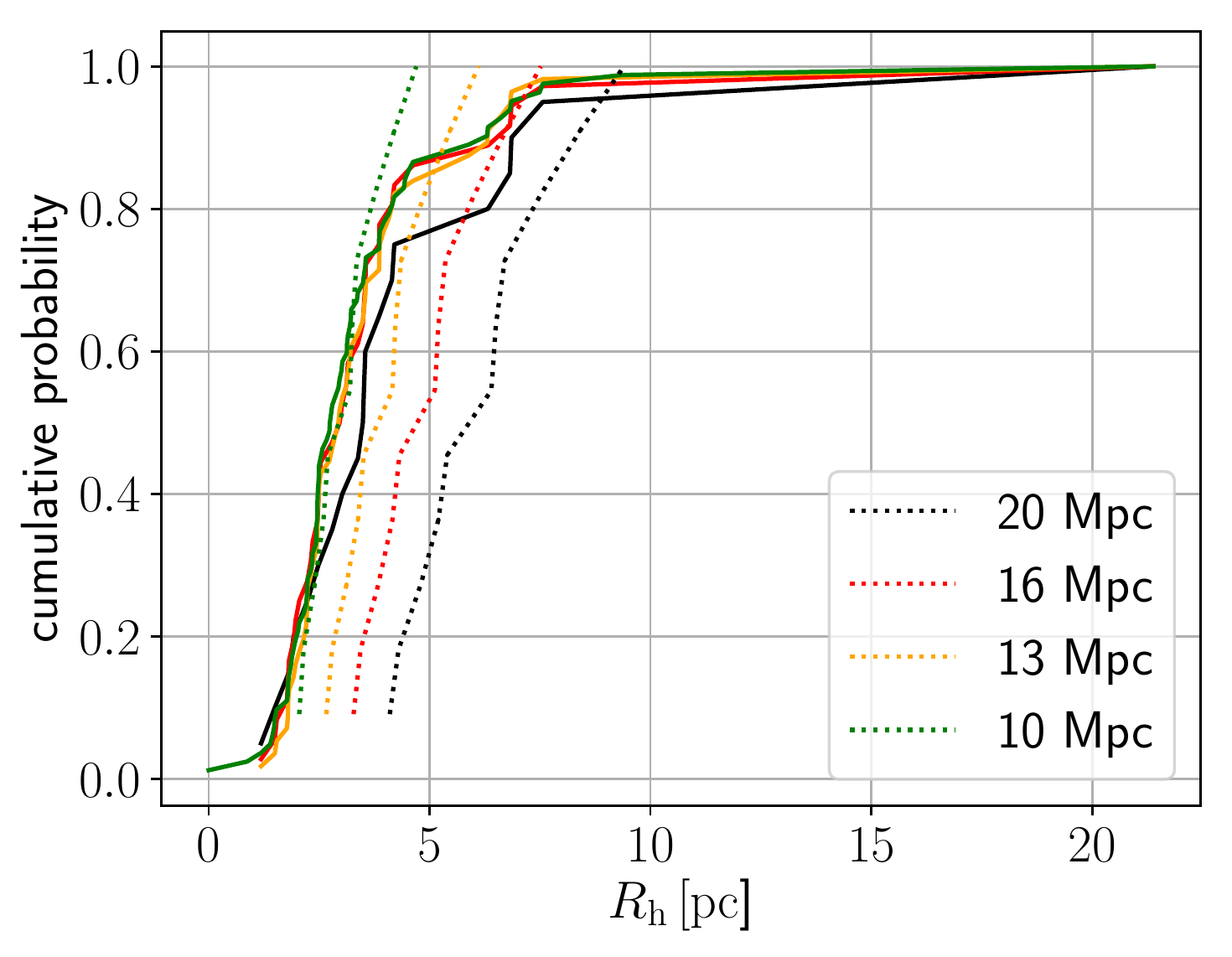}
		\caption{Cumulative probability distributions of the GC half-light radius distributions in the MW (solid lines) and DF2 (dashed lines). Different line colours refer to different distances of the observed DF2 galaxy from Earth (from left to right: $D = 10 \, \rm{Mpc}$, $13 \, \rm{Mpc}$, $16 \, \rm{Mpc}$, and $20 \, \rm{Mpc}$). The maximal differences between these distributions are $0.21$ ($2.0 \, \rm{pc}$), $0.43$ ($2.5 \, \rm{pc}$), $0.58$ ($3.2 \, \rm{pc}$), and $0.66$ ($4.2 \, \rm{pc}$), respectively.}
		\label{fig:KS_test_radii}
	\end{figure}

	The cumulative probability distributions of the DF2 and MW GC half-light radius distributions and their maximum differences, $D_{\mathrm{max}}$, are shown in Fig.~\ref{fig:KS_test_radii}. 
	
	\subsection{GC probabilities}
	Summing up, all the here applied statistical tests have shown that scaling the GC luminosities and sizes of the DF2 dwarf galaxy to $D \approx 10 - 13 \,\rm{Mpc}$ resolves the tension with the observed GC population of the MW which is known to be representative \citep[canonical;][]{Rejkuba_2012}. At $D = 20 \, \rm{Mpc}$ the GC luminosity and half-light radius distributions of DF2 are in significant conflict with the GC population of the MW. 
	
	\begin{table*}
		\caption{$P$-values of the binomial cumulative functions (see method 1 in 
			Section~\ref{sec:Results_Radii_Binomial}), comparison of the means (see Section~\ref{sec:Results_Radii_means}), Mann--Whitney U test (see Section~\ref{sec:Results_Radii_Utest}), and KS test (see Section~\ref{sec:Results_Radii_KStest}) applied on the half-light radius distribution of the GC population in the MW and DF2.}
		\label{tab:results_statistics_radii}
		\begin{tabular}{llllll} \hline 
			$D \, [\rm{Mpc}]$ & $M_{V,606}$ & $P$-value (binomial 1) & $P$-value (mean comp.) & $P$-value (U-test)  & $P$-value (KS test)  \\ \hline \hline 
			$20$ & $\leq -8.6$ & $3.1 \times 10^{-5}$ & $2.0 \times 10^{-1}$ & $6.0 \times 10^{-2}$ & $1.9 \times 10^{-3}$ \\ 
			$16$ & $\leq -8.1$ & $3.1 \times 10^{-5}$ & $2.2 \times 10^{-1}$ & $2.7 \times 10^{-3}$ & $3.4 \times 10^{-3}$ \\ 
			$13$ & $\leq -7.7$ & $4.9 \times 10^{-4}$ & $6.1 \times 10^{-1}$ & $3.8 \times 10^{-2}$ & $4.8 \times 10^{-2}$ \\ 
			$10$ & $\leq -7.1$ & $1.5 \times 10^{-1}$ & $7.5 \times 10^{-1}$ & $6.0 \times 10^{-1}$ & $7.5 \times 10^{-1}$  \\  \hline 
		\end{tabular}
	\end{table*}
	
	\subsection{Combined cosmological and GC probability} \label{sec:combined_probability}
\begin{figure}
		\centering
		\includegraphics[width=\linewidth]{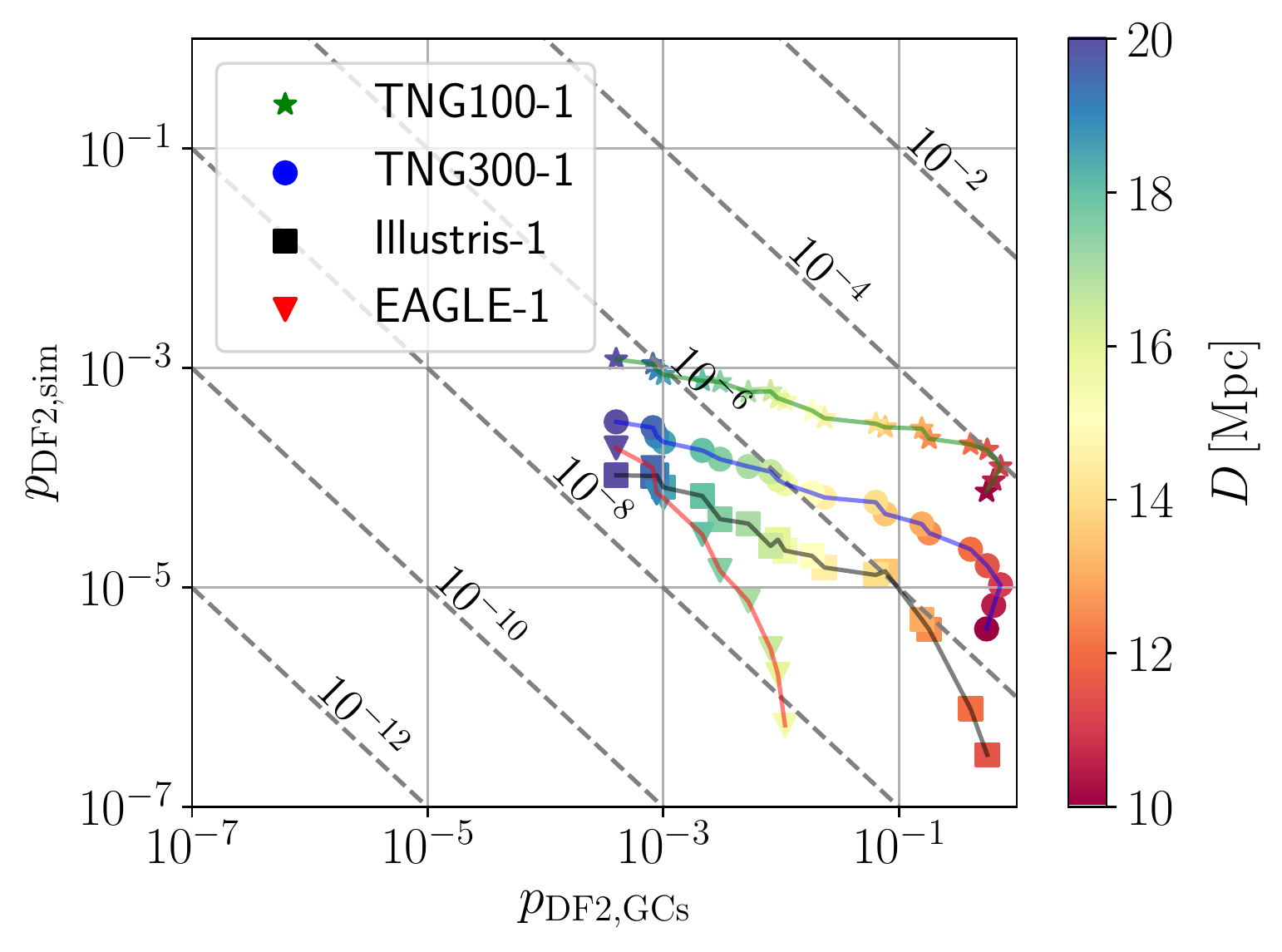}
		\caption{Cosmological detection probabilities of DF2-like dwarf galaxies obtained from all simulation runs, $p_{\mathrm{DF2,sim}}$, (see also Table~\ref{tab:results_all_distances}) and $P$-values of the KS test on the luminosity distribution of the observed DF2 galaxy, $p_{\mathrm{DF2,GCs}}$, (see last column in Table~\ref{tab:results_statistics_luminosities}) for different distances, $D$, represented by the colourbar. The dashed lines mark the positions of constant combined probability. The four different tracks belong to different simulations: Illustris-1 (squares, black line), TNG100-1 (stars, green line), TNG300-1 (dots, blue line), and EAGLE-1 (triangles, red line; see also the different sets in Fig.~\ref{fig:Pcomb2}).}
		\label{fig:Pcomb1}
	\end{figure}

	In Section~\ref{sec:Results_cosmological_simulations} we have shown that given its $M_{\mathrm{total}}/M_{*}$ value, stellar mass and half-mass radius, peculiar velocity, and motion relative to the LG, $\Lambda$CDM simulations favor that DF2 is located at a distance greater than $20 \, \rm{Mpc}$. In contrast, the above statistical tests applied on the GC population of DF2 in Sections~\ref{sec:Results_Luminosities} and \ref{sec:Results_Radii} suggest much smaller distances of around $10 \, \rm{Mpc}$. Correlating in Fig.~\ref{fig:Pcomb1} the probabilities obtained from the cosmological simulations, $p_{\mathrm{DF2,sim}}$, with the $P$-values of the KS test on the GC luminosity distribution (see Section~\ref{sec:Results_Luminosities_KStest}) yields that the existence of such a dwarf galaxy is exceedingly rare in standard cosmology. Here we use the $P$-value of the KS test applied on the GC luminosity distribution, because this is statistically more robust than its GC half-light radius distribution compared to other galaxies (see last column of Tables~\ref{tab:results_statistics_luminosities} and \ref{tab:results_statistics_radii}). The probabilities globally peak at around $12 \, \rm{Mpc}$. Distances less than $10 \, \rm{Mpc}$ are disfavoured by both the luminosity function and the other properties of DF2. This justifies our decision to not consider $D < 10 \, \rm{Mpc}$.
	
\begin{figure}
		\centering
		\includegraphics[width=85mm]{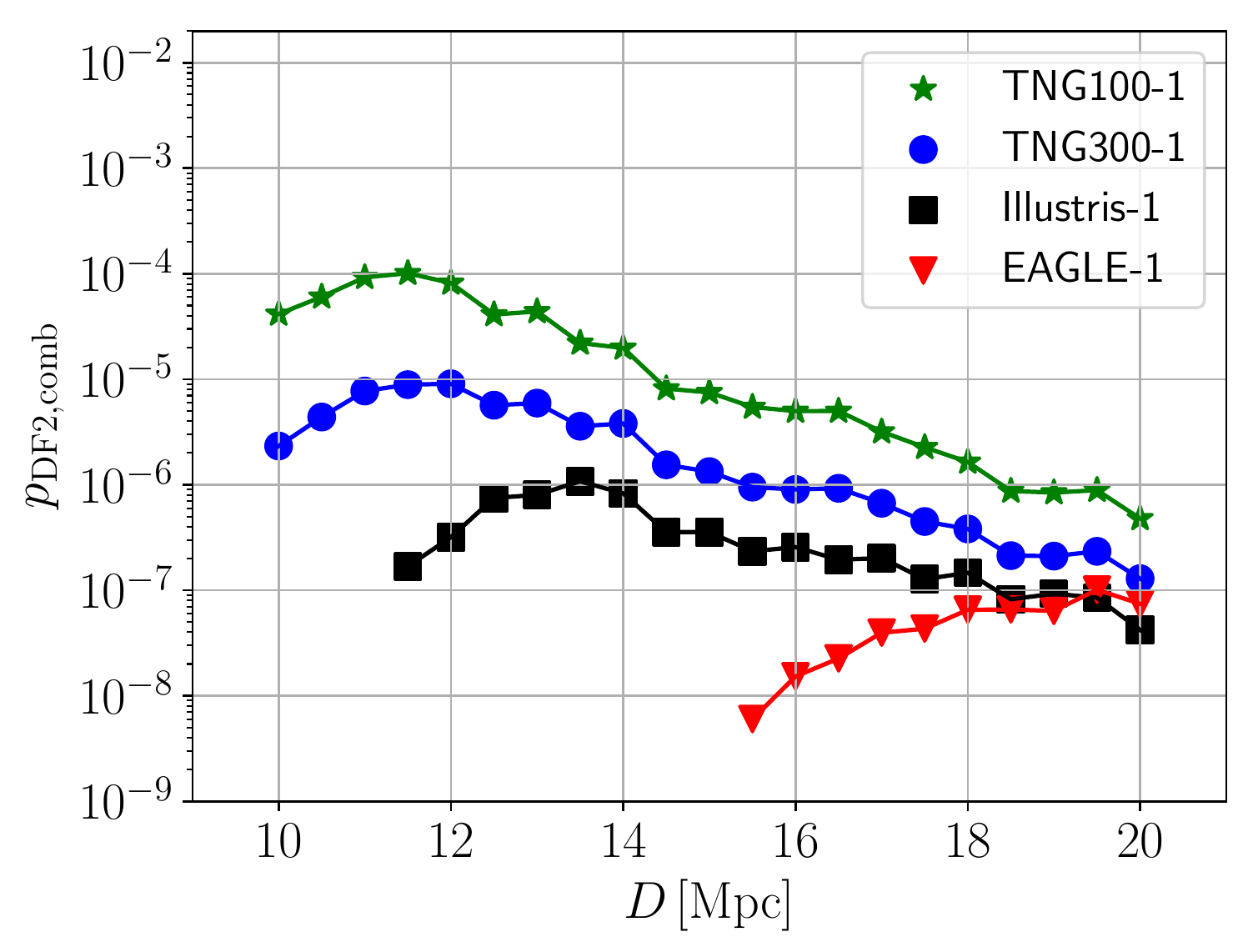}
		\caption{Combined probability of a DF2-like galaxy in standard cosmology, $p_{\mathrm{DF2,comb}}$, in dependence of its distance, $D$, from Earth for the Illustris-1 (black squares), TNG100-1 (green stars), TNG300-1 (blue dots), and EAGLE-1 (red triangles) simulations (see also Table~\ref{tab:results_combined} for the absolute numbers). There are no DF2-like subhaloes in the EAGLE-2 and EAGLE-3 simulations. The combined probability, $p_{\mathrm{DF2,comb}}$, is calculated by multiplying the probabilities from the $\Lambda$CDM simulations, $p_{\mathrm{DF2,sim}}$, (see Table~\ref{tab:results_all_distances}) with the $P$-values of the KS test on the GC luminosity populations. 
			}
		\label{fig:Pcomb2}
	\end{figure}

	In Fig.~\ref{fig:Pcomb2} and Table~\ref{tab:results_combined} the combined probability, $p_{\mathrm{DF2,comb}}$, is calculated by multiplying the probabilities from cosmological simulations, $p_{\mathrm{DF2,sim}}$, with the $P$-values of the KS test by assuming that the GC population is independent of the structural properties of galaxies. The combined probability is maximal in the TNG100-1 simulation and peaks at a distance of $11.5 \pm 1.5 \, \rm{Mpc}$ with $ 1.0 \times 10^{-4}$ making the observed DF2 dwarf galaxy very unlikely in standard cosmology. This global peak is mainly driven by the GC population but is also apparent based on the structural properties (left-hand panel of Fig. \ref{fig:Pstructure}).
	
	NGC 1052 has a stellar mass of $M_{*} \approx 10^{11} \, \rm{M_{\odot}}$ \citep{Forbes_2017}, which implies a host halo mass of $M_{200} \approx 10^{13} \, \rm{M_{\odot}}$, and has a projected separation of $80 \, \rm{kpc}$ from DF2 if located at $D = 20 \, \rm{Mpc}$. In Appendix~\ref{appendix:combined_probability} we show that restricting the analysis on DF2-like subhaloes embedded in host halos with $M_{200} \leq 10^{13} \, \rm{M_{\odot}}$ reduces the probabilities of DF2-like galaxies in the Illustris-1, TNG100-1, and TNG300-1 simulations by about one order of magnitude. DF2 then would be a $4.6 \sigma$ outlier within the $\Lambda$CDM framework.  
	
	\begin{table*}
		\caption{Combined probabilities, $p_{\mathrm{DF2,comb}}$, of the existence of a DF2-like galaxy based on cosmological simulations and its GC population using the $P$-values of the KS test applied on the GC luminosity distribution (see Section~\ref{sec:Results_Luminosities_KStest} and last column of Table~\ref{tab:results_statistics_luminosities}) for different distances (see also Figs~\ref{fig:Pcomb1} and \ref{fig:Pcomb2}).}
		\label{tab:results_combined}
		\begin{tabular}{lllllll} \hline 
			Distances & $20 \, \rm{Mpc}$ \citep{vDokkum_2018a} & $20 \, \rm{Mpc}$ \citep{Martin_2018} & $16 \, \rm{Mpc}$ & $13 \, \rm{Mpc}$ & $10 \, \rm{Mpc}$ \\ \hline  
			Simulation        & $p_{\mathrm{DF2,comb}}$  & $p_{\mathrm{DF2,comb}}$ & $p_{\mathrm{DF2,comb}}$ & $p_{\mathrm{DF2,comb}}$ & $p_{\mathrm{DF2,comb}}$  \\ \hline \hline  
			Illustris-1       & $1.6 \times 10^{-8}$  & $4.2 \times 10^{-8}$ & $2.6 \times 10^{-7}$ & $8.0 \times 10^{-7}$ & $0$ \\  
			TNG100-1 & $3.6 \times 10^{-8}$  & $4.8 \times 10^{-7}$ & $5.0 \times 10^{-6}$ & $4.4 \times 10^{-5}$ & $4.2 \times 10^{-5}$ \\  
			TNG300-1 & $9.9 \times 10^{-9}$  & $1.3 \times 10^{-7}$ & $9.0 \times 10^{-7}$ & $5.9 \times 10^{-6}$ & $2.3 \times 10^{-6}$ \\  
			EAGLE-1           & $0$ & $7.4 \times 10^{-8}$  & $1.5 \times 10^{-8}$ & $0$ & $0$\\  
			EAGLE-2           & $0$  & $0$ & $0$ & $0$ & $0$ \\  
			EAGLE-3           & $0$  & $0$ & $0$ & $0$ & $0$ \\ 
			\hline 
		\end{tabular}
	\end{table*}
	
	\section{Discussion} \label{sec:Discussion}
	
	Assuming that the standard model of cosmology is a correct description of the observed Universe, both dark-matter-dominated dwarf galaxies and TDGs void of dark matter must exit \citep{Kroupa_2012,Haslbauer_2019}. Therefore the discovery of the two dark-matter-deficient galaxies DF2 and DF4 by \citet{vDokkum_2018a, vDokkum_2019} are in principle consistent with the $\Lambda$CDM paradigm. Recently, \citet{Ploeckinger_2018} and \citet{Haslbauer_2019} reported that dark-matter-free dwarf galaxies (i.e. tidal dwarf galaxy candidates) can be found in self-consistent cosmological simulations. However, based on the occurrence in cosmological simulations and the GC population of DF2, we have shown in this work that the chance of finding similar galaxies in standard cosmology is extremely low, with a maximal probability of $1.0 \times 10^{-4}$ at a distance of $D = 11.5\pm1.5 \, \rm{Mpc}$ obtained from the TNG100-1 simulation and assuming an $M/L$ ratio consistent with \citet{Martin_2018} (see Section~\ref{sec:combined_probability}). It would be $2.4 \times 10^{-5}$ at $D=11.0\pm0.5 \, \rm{Mpc}$ for the $M/L$ ratio of \citet{vDokkum_2018a}.
	
	The probabilities in cosmological simulations are significantly reduced further if one focuses only on dark-matter-lacking galaxies embedded in host halos with $M_{200} \leq 10^{13} \, \rm{M_{\odot}}$ motivated by NGC 1052, which is in close location to DF2. This implies that such simulated galaxies are typically found in very massive galaxy clusters. Again, the highest probability is achieved in the TNG100-1 simulation with $5.2 \times 10^{-6}$ at a slightly larger distance of $13.0\pm1.5 \, \rm{Mpc}$ (see Appendix~\ref{appendix:combined_probability}). Such a distance is consistent with the work of \citet{Trujillo_2019} in which they conclude that DF2 is located at $D = 13.0 \pm 0.4 \, \rm{Mpc}$ based on redshift-independent distance indicators \citep[see also fig.~18  in][]{Trujillo_2019}.
	
	In particular, without considering the unusual GC population, cosmological simulations suggest that DF2 is most likely located at a distance of $D > 20 \, \rm{Mpc}$ from Earth. The probabilities of DF2-like galaxies with a dark matter content estimated from \citet{vDokkum_2018a} and \citet{Danieli_2019} ($M/L < 2  \, \rm{\Upsilon_{\odot}}$) are about a factor of $10 \, \times$ smaller than for galaxies with a slightly higher amount of dark matter as proposed by \citet{Martin_2018} ($M/L < 8.1  \, \rm{\Upsilon_{\odot}}$). Thus, in order to minimize the tensions with $\Lambda$CDM cosmology, the analysis of the frequency of DF2 analogues over distance is taken here to rely on the description of \citet{Martin_2018}. In the following we discuss the role of the structural properties and peculiar velocity on the frequency of DF2-like subhaloes. 
	
	\subsection{Comparison of simulations} \label{sec:Comparison of simulations}

	\subsubsection{Structural properties of subhaloes} \label{sec:Structural properties of subhaloes}
	According to the Dual Dwarf Theorem dark-matter-deficient galaxies are more compact than dark-matter-dominated galaxies \citep{Kroupa_2010, Dabringhausen_2013, Haslbauer_2019}. The observed DF2 galaxy is located, in the radius-mass-diagrams of the Illustris-1 simulation, almost exactly between dark-matter-poor and dark-matter-dominated galaxy branches \citep{Haslbauer_2019}. That this cosmological simulation cannot reproduce the correct galaxy sizes of dark-matter-dominated galaxies (compared to observed galaxies) causes in the first place the low detection probabilities of DF2-like objects in the Illustris-1 simulation. The untypical stellar half-mass radii of simulated galaxies could be related to the implemented feedback models in the here analysed simulations. We have used the most modern self-consistent cosmological simulations, which rely on different feedback descriptions, galaxy evolution models, and different computer codes. The higher probabilities (Tables~\ref{tab:results_all_distances} and \ref{tab:results_20Mpc}) of DF2-like galaxies in TNG100-1 and TNG300-1 can be explained by the fact that the TNG simulation produces stellar half-mass radii about $2 \, \times$ smaller than in the Illustris simulations for galaxies with $M_{*} < 10^{10} \, \rm{M_{\odot}}$ \citep{Pillepich_2018}. 
	
\begin{figure*}
		\centering
		\includegraphics[width=88mm]{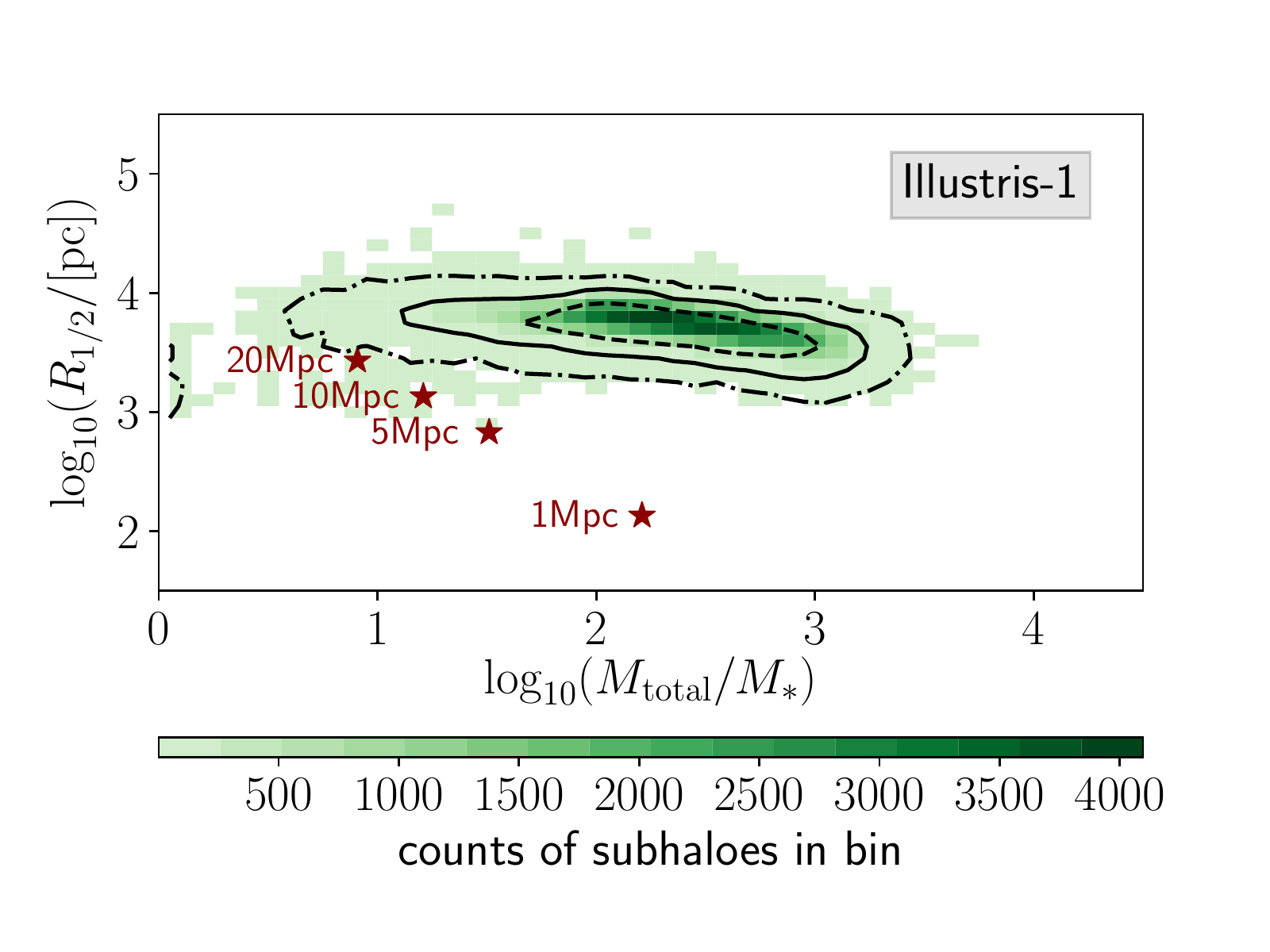}
		\includegraphics[width=88mm]{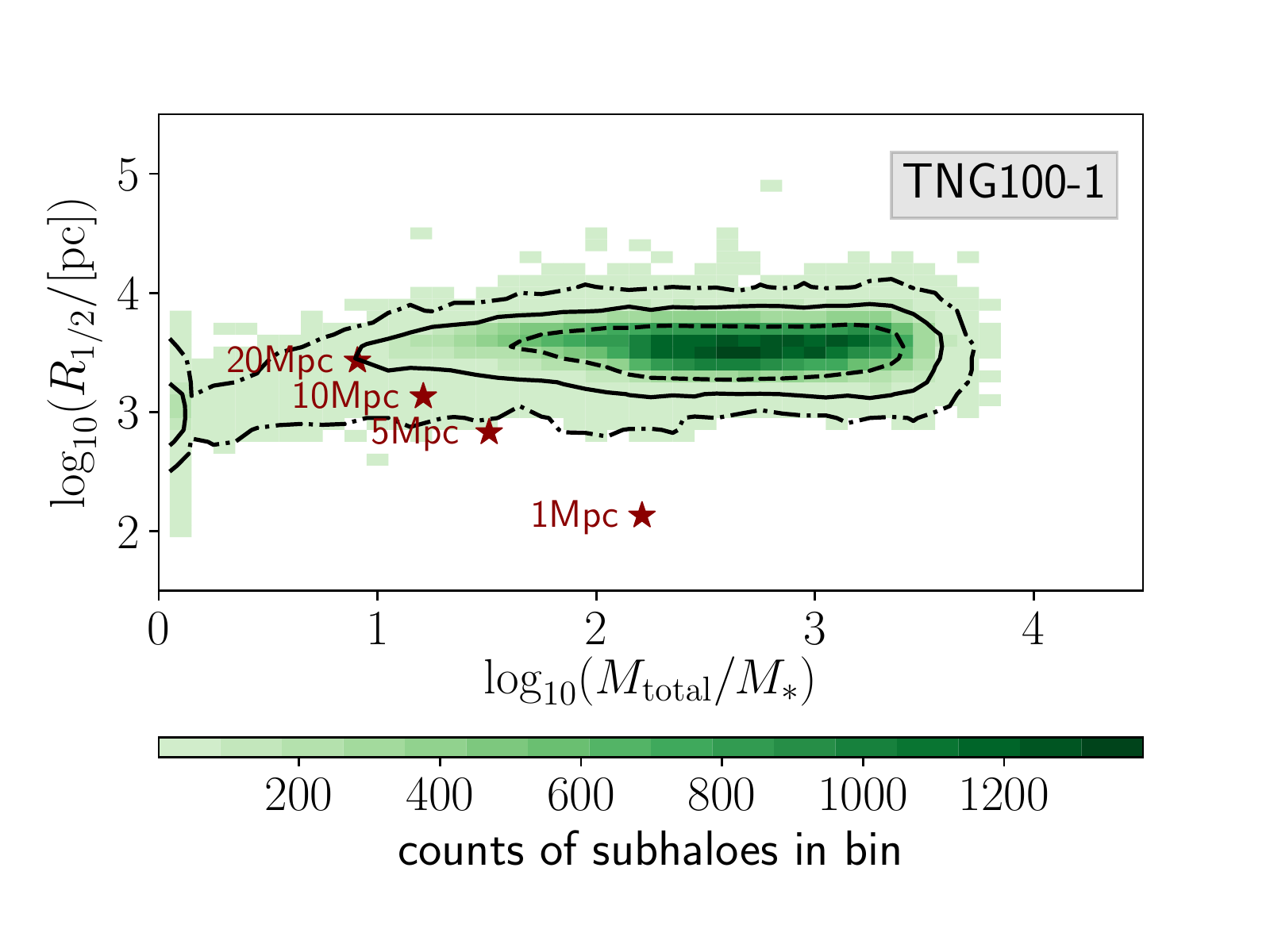}
		
		\includegraphics[width=88mm]{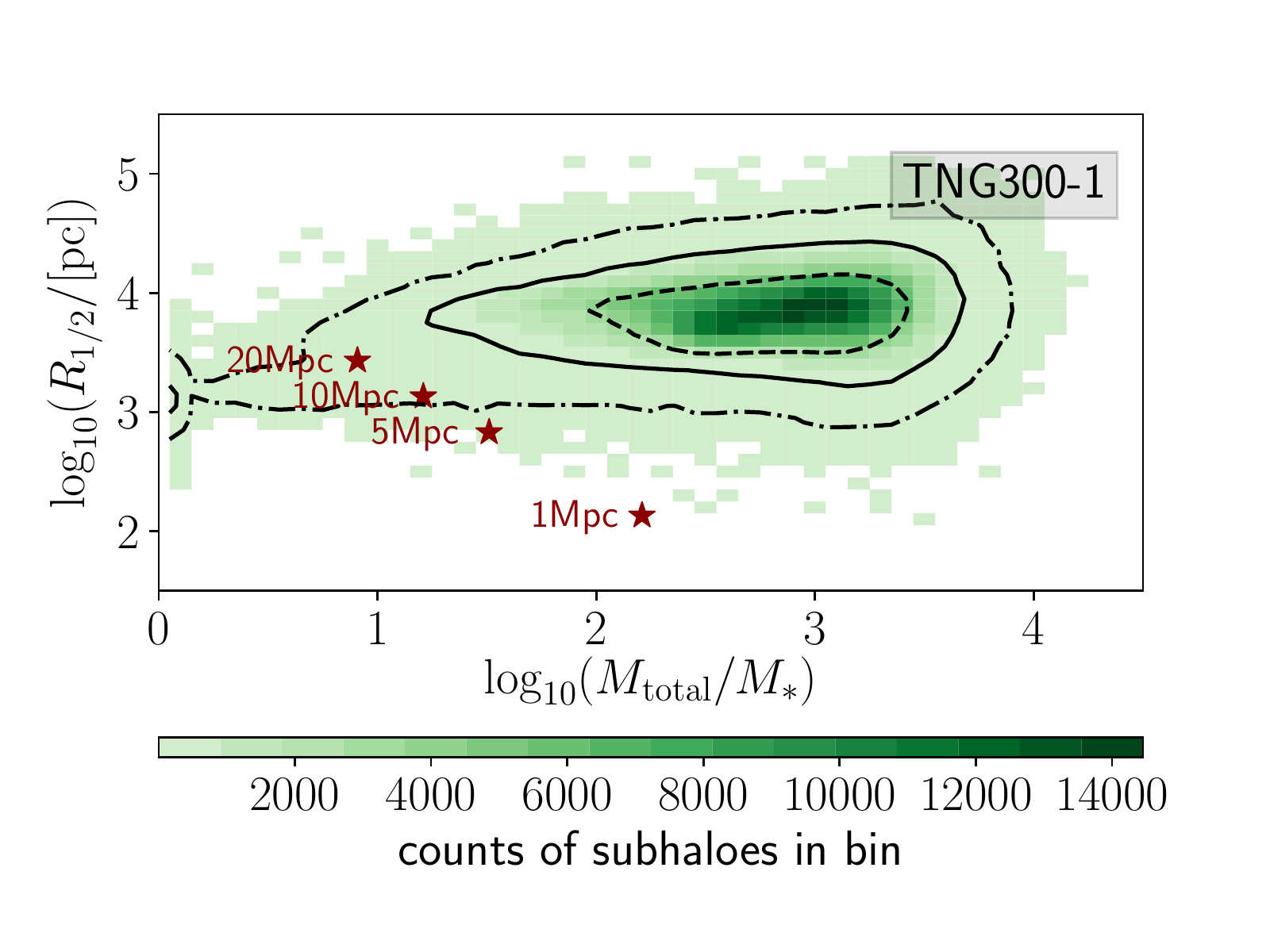}
		\includegraphics[width=88mm]{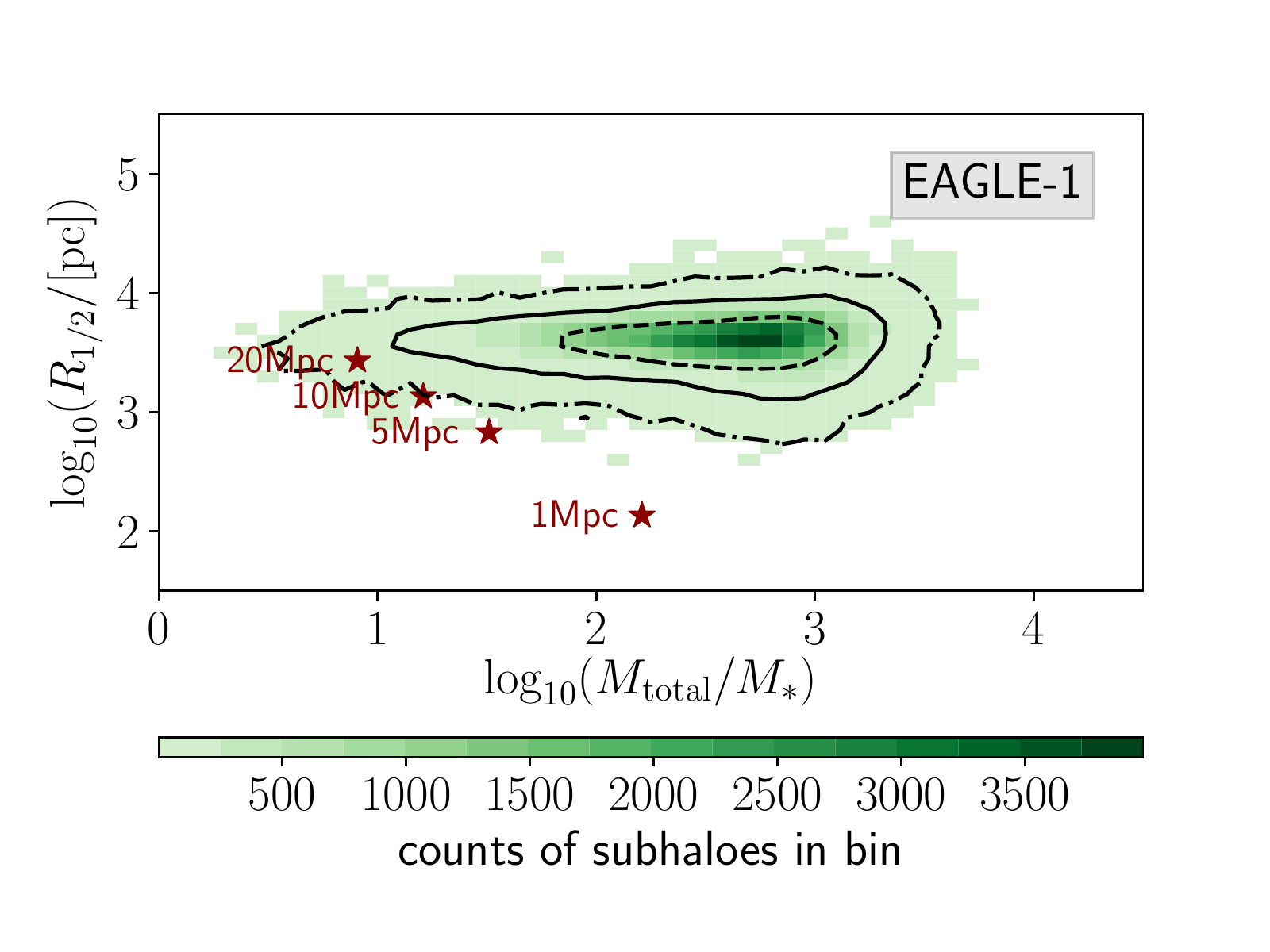}
		
		\includegraphics[width=88mm]{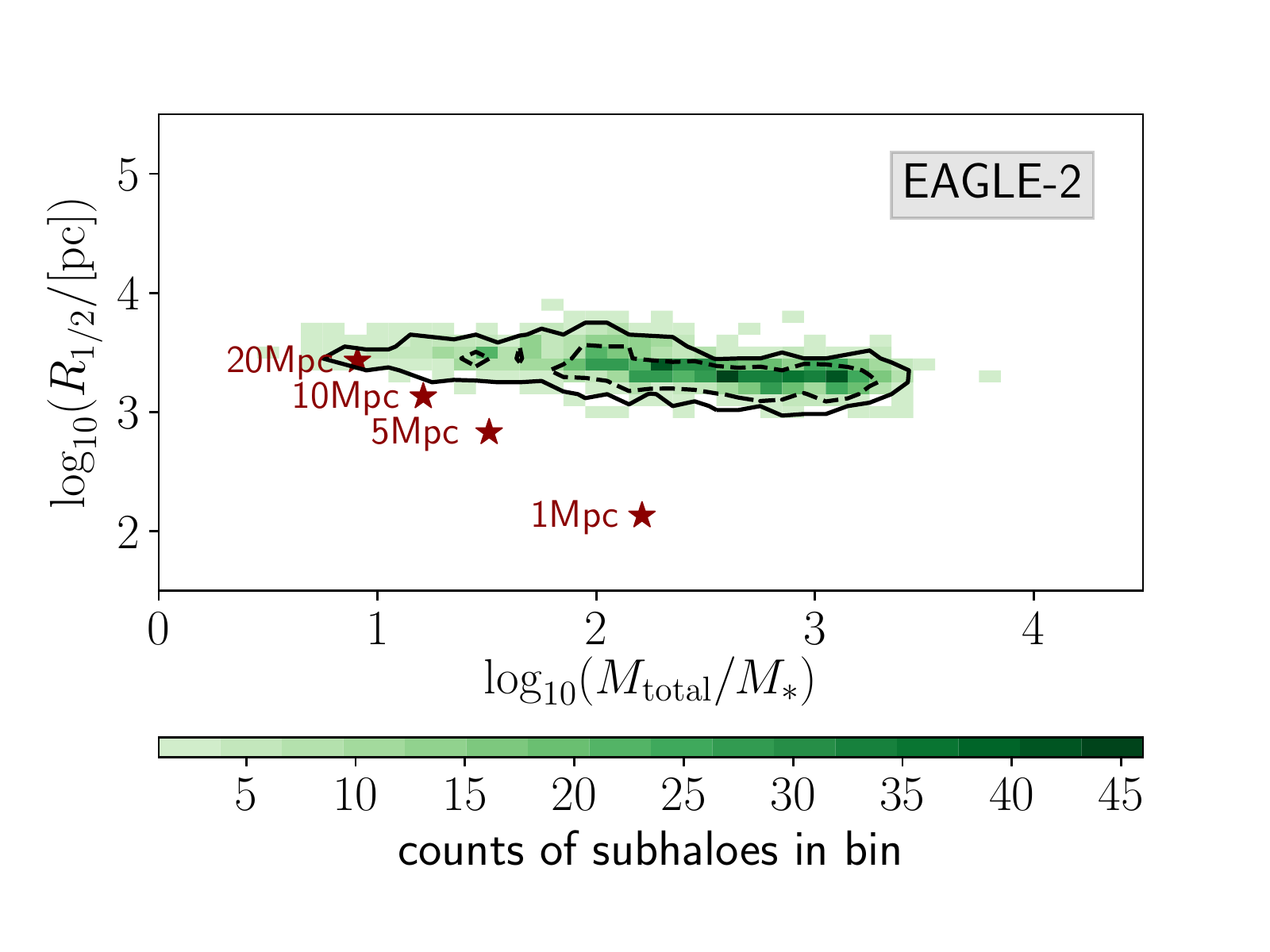}
		\includegraphics[width=88mm]{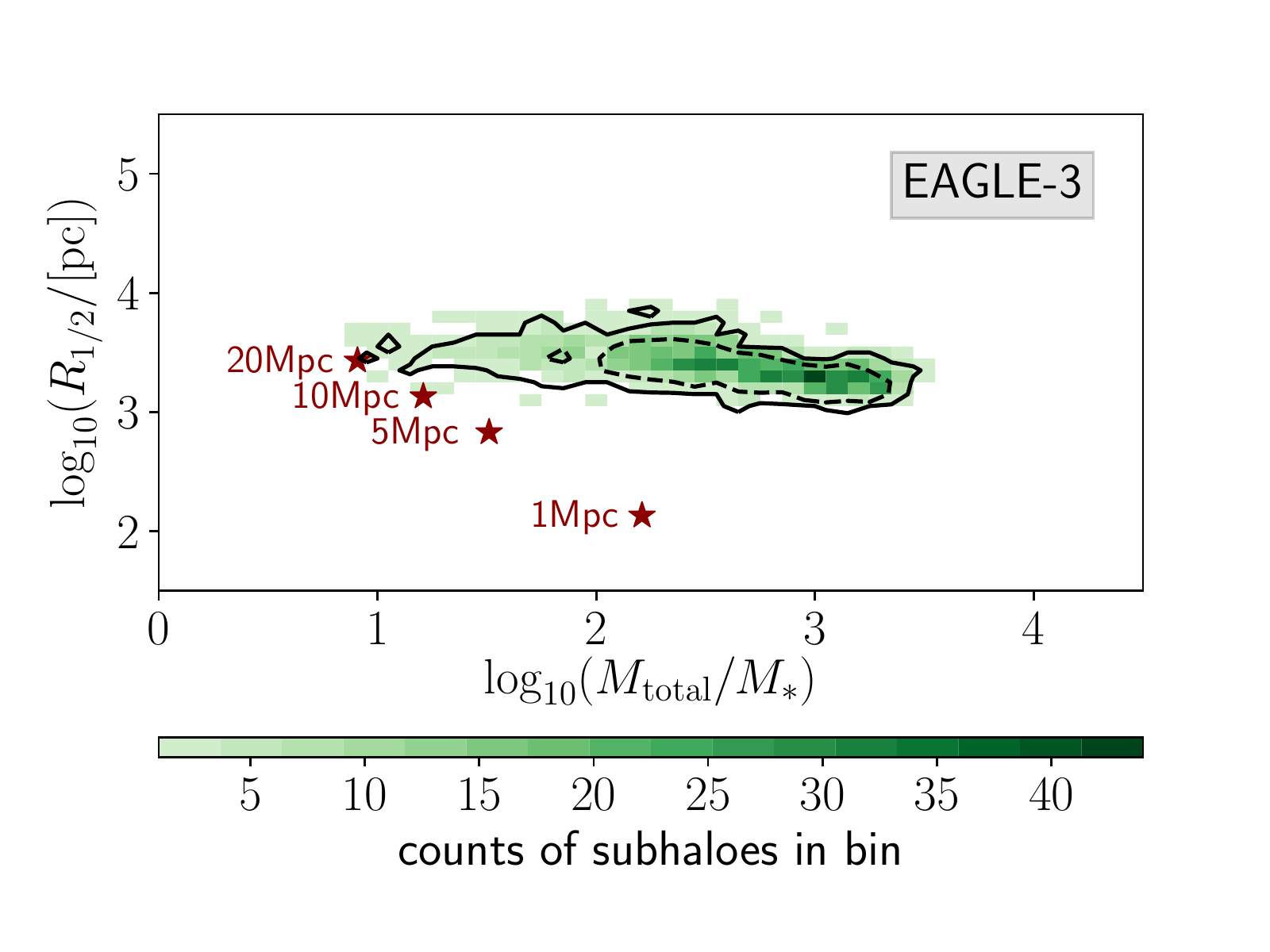}
		
		\caption{Radius--$M_{\mathrm{total}}/M_{*}$ diagram for substructure-corrected subhaloes with $10^{7} < M_{*} / \rm{M_{\odot}} <  10^{9}$ in the Illustris-1, TNG100-1, TNG300-1, EAGLE-1, EAGLE-2, and EAGLE-3 simulations. The distance-dependent probability density of simulated DF2 analogues (highlighted by red stars) is analysed in Fig.~\ref{fig:structure2_simulated_DF2}. The black lines mark the $1\sigma$ (dashed), $2 \sigma$ (solid), and $3\sigma$ (long-dashed) confidence levels if shown. Using the contour plot of the TNG100-1 simulation (upper right panel), the structural properties of DF2 are in tension with $\Lambda$CDM cosmology at the $2.0 \sigma$ ($D = 20 \, \rm{Mpc}$), $2.1 \sigma$ ($D = 16 \, \rm{Mpc}$), $2.3 \sigma$ ($D = 13 \, \rm{Mpc}$), and $2.5\sigma$ ($D = 10 \, \rm{Mpc}$) level.}
		\label{fig:structure1_simulated_DF2}
	\end{figure*}

\begin{figure}
		\centering
		\includegraphics[width=85mm]{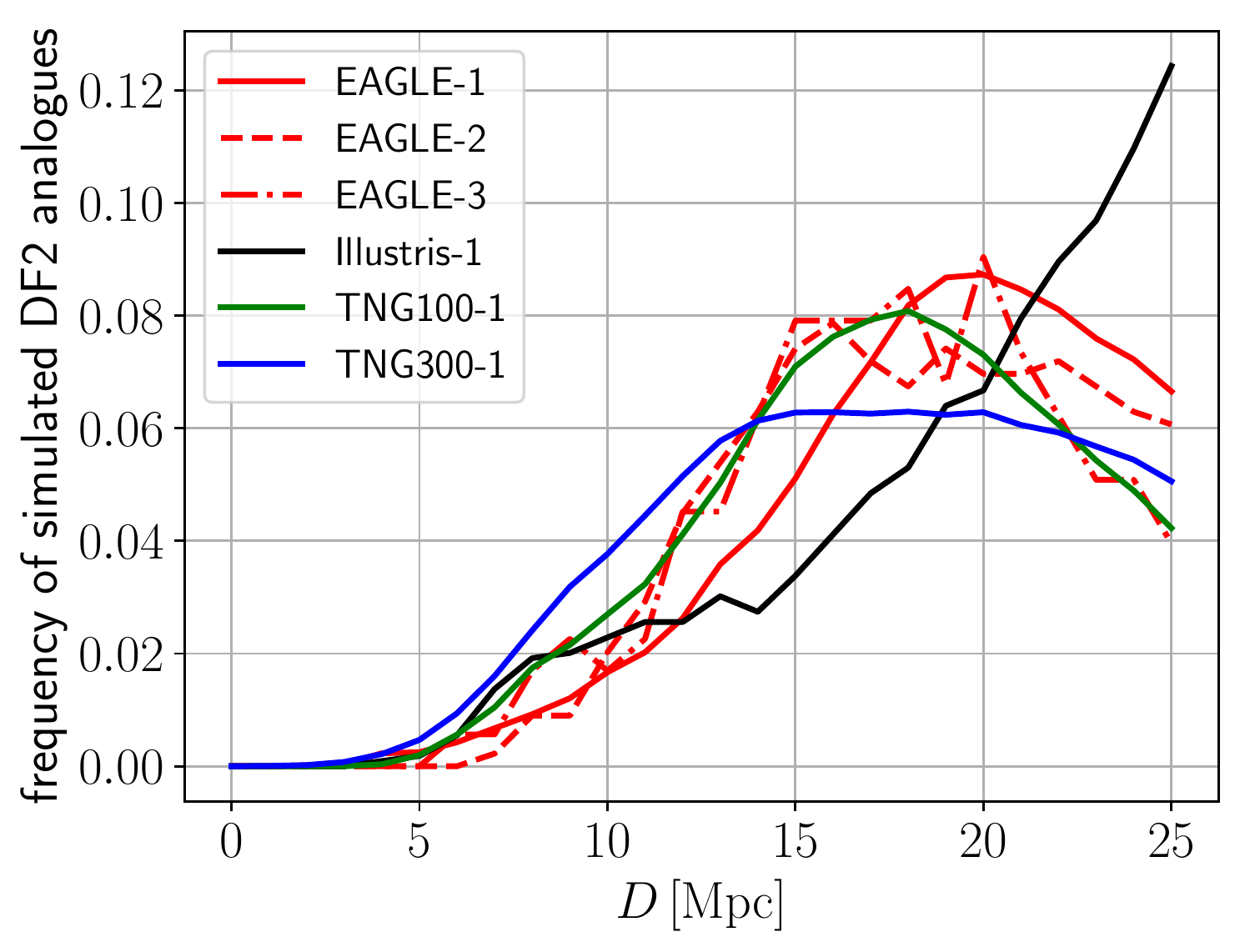}
		\caption{Distance-dependent probability density of DF2 analogues in the Illustris-1 (black), TNG100-1 (green), TNG300-1 (blue), and EAGLE (red) simulations with $10^{7} < M_{*} / \rm{M_{\odot}} <  10^{9}$ calculated by extracting all bin values along the locus of possible DF2 positions in the radius--$M_{\mathrm{total}}/M_{*}$ diagrams of Fig.~\ref{fig:structure1_simulated_DF2} (highlighted by red stars).}
		\label{fig:structure2_simulated_DF2}
	\end{figure}

	The sizes of subhaloes are studied in Fig.~\ref{fig:structure1_simulated_DF2} in which the radius-$M_{\mathrm{total}}/M_{*}$ diagrams for substructure-corrected subhaloes with $10^{7} <M_{*}/\rm{M_{\odot}}< 10^{9}$ in the here used simulation runs and the position of DF2 scaled for different distances (red stars) are plotted. Extracting the bin values along this locus of possible DF2 positions gives a distance-dependent probability distribution as shown in Fig.~\ref{fig:structure2_simulated_DF2}. The frequency of simulated DF2 analogues in the TNG100-1, TNG300-1, and EAGLE runs peaks between $D= 13 - 20 \, \rm{Mpc}$. In contrast, the frequency of Illustris-1 DF2 analogues increases for larger distances because subhaloes in the Illustris-1 simulation are typically less compact than in the other simulations and scaling DF2 to larger distances increases its half-light radius (see also the scaling relations in Appendix~\ref{appendix:scaling_relations}). According to the TNG100-1 radius--$M_{\mathrm{total}}/M_{*}$ contour diagram (upper right panel of Fig.~\ref{fig:structure1_simulated_DF2}), which does not consider the peculiar velocity and GC population, DF2 is at a $2.0 \sigma$ ($D = 20 \, \rm{Mpc}$), $2.1 \sigma$ ($D = 16 \, \rm{Mpc}$), $2.3 \sigma$ ($D = 13 \, \rm{Mpc}$), and $2.5\sigma$ ($D = 10 \, \rm{Mpc}$) tension level with $\Lambda$CDM cosmology. This method, which quantifies the tension with $\Lambda$CDM by using contour plots, differs from that of our main analysis in which we calculated the structural probabilities by using the area of a rectangular defined by $R_{1/2}$ and $M_{\mathrm{total}}/M_{*}$ scaled for different distances (see Table~\ref{tab:selectioncriteria}). The so-calculated structural probability of DF2-like subhaloes in dependence of its distance (see left-hand panel of Fig.~\ref{fig:Pstructure}) shows that the probability becomes maximal for the TNG100-1 simulation. Applying selection criteria for $R_{1/2}$, $M_{*}$, and $M_{\mathrm{total}}/M_{*}$ causes a decrease in the number of subhaloes by about two orders of magnitude. The frequency of such DF2-like subhaloes is about $10\times$ lower in the TNG300-1 simulation compared to TNG100-1. This can be caused either by the higher resolution of TNG100-1 or it means that the frequency of selected subhaloes in TNG100-1 has not converged since the box size of TNG100-1 is by a factor of $3 \times$ smaller than TNG300-1.
	
	Using the Illustris-1 \citet{Yu_2018} find a large population of simulated dwarf galaxies with $M_{*} < 5 \times 10^{8} \, \rm{M_{\odot}}$ and $M_{\mathrm{dm}}/M_{*} <~0.1$ (see their Fig.~3). However, their analysis does not include any constraints on the sizes, environment, and peculiar velocity of dark-matter-lacking subhaloes. The role of the peculiar velocity is discussed in the following section. 
	
	\subsubsection{Velocity field of subhaloes} \label{sec:Velocity field of subhaloes}
	The analysis of simulated DF2 analogues is based on simulation runs with different computational box sizes. Small simulation boxes have a lack of large-scale waves, which affects the peculiar velocities in two ways. First, cosmological simulations do not have bulk velocities relative to the CMB. Secondly, the absence of large-scale waves causes a lack of large-scale fluctuations and thus massive clusters. Therefore, peculiar velocities of subhaloes in small simulation boxes, like the Illustris-1, TNG100-1, and EAGLE runs, are not representative of a ``real'' $\Lambda$CDM Universe. 
	
	In order to address the lack of large-scale waves/fluctuations in cosmological simulations, we also used the TNG300-1 simulation which is the largest run of the IllustrisTNG project with a co-moving box size of $L = 302.6 \, \rm{cMpc}$. The LG moves with a peculiar velocity of about $627 \, \rm{km \, s^{-1}}$ wrt. the CMB rest frame, which is most likely caused by the Shapely supercluster and the GA. Assuming that a large-scale structure causes a peculiar velocity of $v_{\mathrm{pec}} = 627 \, \rm{km \, s^{-1}}$ in the region of the LG, we estimate its maximal induced peculiar velocity $\Delta v_{\mathrm{pec}}$ on the motion of DF2 if such a supposed large-scale structure is just outside of the simulation box by
	
	\begin{eqnarray}
		\Delta v_{\mathrm{pec}} &=& \bigg( \frac{2 D}{L} \bigg) v_{\mathrm{pec}} \, .
		\label{eq:velocityfield}
	\end{eqnarray}
	The factor $2/L$ takes into account that the LG-like structure is located in the middle of the simulation box and that the large-scale structure is exactly at the edge of the simulation box, which would be the worst situation for our analysis since it maximizes $\Delta v_{\mathrm{pec}}$. $D$ is again the distance between the LG and DF2.
	
	The maximum induced peculiar velocity raised by the tides of a massive large-scale structure outside the simulation box would be rather small for the TNG300-1 simulation (i.e. $\Delta v_{\mathrm{pec}} = 54 \, \rm{km \, s^{-1}}$ for $D = 13 \, \rm{Mpc}$ and $\Delta v_{\mathrm{pec}} = 83 \, \rm{km \, s^{-1}}$ for $D = 20 \, \rm{Mpc}$), but is significantly larger for the Illustris-1, TNG100-1, and EAGLE simulations (i.e. for the TNG100-1 simulation one gets: $\Delta v_{\mathrm{pec}} = 147 \, \rm{km \, s^{-1}}$ for $D = 13 \, \rm{Mpc}$ and $\Delta v_{\mathrm{pec}} = 227 \, \rm{km \, s^{-1}}$ for $D = 20 \, \rm{Mpc}$). 

\begin{figure}
		\centering
		\includegraphics[width=\linewidth]{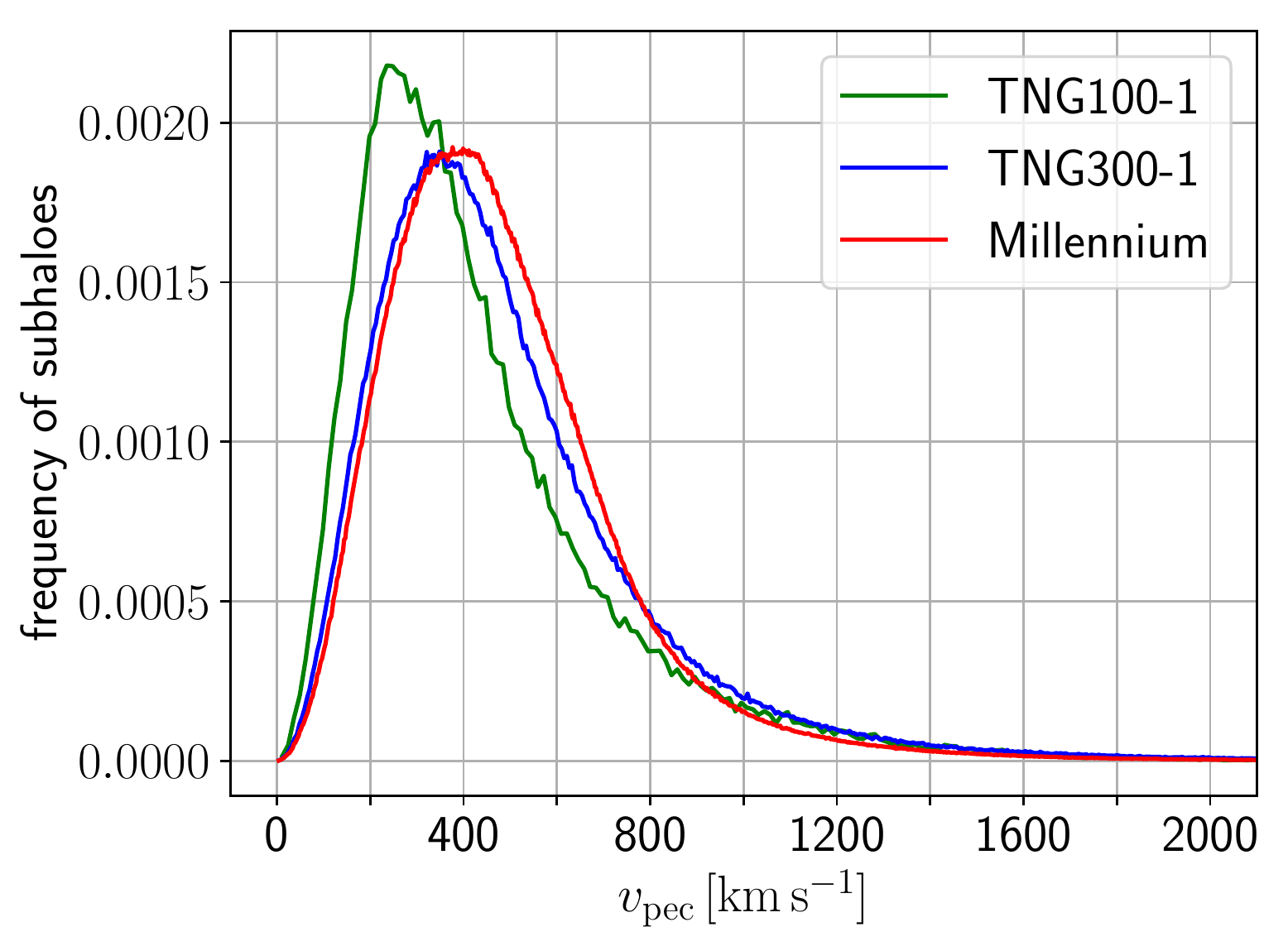}
		\caption{Peculiar velocity field at redshift $z = 0$ of subhaloes with $M_{*} > 10^{7} \, \rm{M_{\odot}}$ in the TNG100-1 (green; $L = 110.7 \, \rm{cMpc}$), TNG300-1 (blue; $L = 302.6 \, \rm{cMpc}$), and Millennium simulation (red; $L = 684.9 \, \rm{cMpc}$, \citealt{Springel_2005B, DeLucia_2006}). The distributions for different stellar mass cuts are analysed in more detail in Table~\ref{tab:peculiar_field_statistics}.}
		\label{fig:velocity_field_LCDM_simulations}
	\end{figure}

	Since the simulation box of TNG300-1 is much larger than those of the other simulations, it includes also much longer wave lengths resulting in a shift of the mode of the peculiar velocity field to higher values as seen in Table~\ref{tab:peculiar_field_statistics} and Fig.~\ref{fig:velocity_field_LCDM_simulations}. In addition, we compare the velocity field of the TNG100-1 and TNG300-1 with the much larger Millennium simulation. The Millennium simulation is an $N$-body simulation carried out in a cosmological box with a length of $L = 684.9 \, \rm{cMpc}$ based on a flat $\Lambda$CDM cosmology with $\Omega_{\mathrm{m}} = 0.25$, $\Omega_{\mathrm{b}} = 0.045$, $\Omega_{\mathrm{\Lambda}} = 0.75$, $h = 0.73$, $n = 1$, and $\sigma_{8} = 0.9$ \citep{Springel_2005B}. Here we use the subhalo catalogue generated with the semi-analytical model from \citet{DeLucia_2006}. Due to the large box size of the Millennium simulation, one can assume that it is a very accurate representation of a $\Lambda$CDM Universe. According to Table~\ref{tab:peculiar_field_statistics} and Fig.~\ref{fig:velocity_field_LCDM_simulations} the velocity field of TNG300-1 is comparable with that of the Millennium simulation, i.e. the modes of the peculiar velocity of subhaloes are very similar for different stellar mass ranges and the standard binomial uncertainties of the velocity fields are very small. Thus, we can conclude that the velocity field has almost converged in the TNG300-1 simulation. Interestingly, the frequency of subhaloes with $v_{\mathrm{pec}} \approx 780 \, \rm{km \, s^{-1}}$ is slightly higher in the TNG300-1 compared to the Millennium simulation. The cross-over between the TNG100-1 and Millennium simulations happens at $v_{\mathrm{pec}} \approx 890 \, \rm{km \, s^{-1}}$. Therefore, TNG300-1 (and also TNG100-1) can be safely used in order to address the high peculiar velocity of DF2 which arises at a small distance. In particular, DF2 has $v_{\mathrm{pec}} = 780 \, \rm{km \, s^{-1}}$ at a distance of $D = 11.5 \, \rm{Mpc}$ from Earth. 
	
	The role of the peculiar velocity on the detection probabilities of DF2-like subhaloes for different simulation runs was studied in Section~\ref{sec:Results_cosmological_simulations} (right-hand panel of Fig.~\ref{fig:Pstructure}). At $D = 20 \, \rm{Mpc}$, the constraints on the peculiar velocity reduce their frequency in the TNG100-1 and TNG300-1 by a factor of $5-10$. This factor increases for smaller distances and at $D = 10 \, \rm{Mpc}$ the peculiar velocity conditions cause a drop in the frequency of at least two orders of magnitude. Interestingly, this effect is almost the same for the TNG100-1 and TNG300-1 simulation. This can be also just a coincidence. 
	Since TNG100-1 has a better resolution than TNG300-1 (see Table~\ref{tab:parameters}) and the fact that the effect on the peculiar velocity is almost the same for both simulations, the main results of our analysis rely on TNG100-1.
		
	\begin{table}
		\centering
		\caption{Fraction of subhaloes with $v_{\mathrm{pec}} \geq 627 \, \rm{km \, s^{-1}}$ and the mode of the $v_{\mathrm{pec}}$ distribution in the TNG100-1 ($L = 110.7 \, \rm{cMpc}$), TNG300-1 ($L = 302.6 \, \rm{cMpc}$), and Millennium simulation ($L = 684.9 \, \rm{cMpc}$, \citealt{Springel_2005B, DeLucia_2006}). We calculate standard binomial uncertainties by $\sqrt{p(1-p)/n}$, where $p$ is the fraction of subhaloes with $v_{\mathrm{pec}} \geq 627 \, \rm{km \, s^{-1}}$ and $n$ is the total number of subhaloes of the distribution. These uncertainties are always smaller than $0.005$. The uncertainties of the velocity peaks are estimated by the chosen bin sizes of the histograms and the velocity fields are shown in Fig.~\ref{fig:velocity_field_LCDM_simulations}. }
		\label{tab:peculiar_field_statistics}
		\begin{tabular}{llllllllllll} \hline 
			$M_{*} \, [\rm{M_{\odot}}]$ & TNG100-1 & TNG300-1 & Millennium \\ \hline \hline 
			\multirow{2}{*}{$> 10^{7}$} & $0.185$ & $0.236$ & $0.220$ \\ 
			& $254 \pm 20 \, \rm{km \, s^{-1}}$ & $351 \pm 20 \, \rm{km \, s^{-1}}$ & $392 \pm 20 \, \rm{km \, s^{-1}}$ \\
			\multirow{2}{*}{$10^{7}-10^{9}$}   & $0.185$ & $0.233$ & $0.203$ \\ 
			& $254 \pm 20 \, \rm{km \, s^{-1}}$ & $351 \pm 20 \, \rm{km \, s^{-1}}$ & $352 \pm 20 \, \rm{km \, s^{-1}}$ \\
			\multirow{2}{*}{$10^{10}-10^{12}$} & $0.185 $ & $0.250$  & $0.275$ \\  
			& $ 255 \pm 20 \, \rm{km \, s^{-1}}$ & $355 \pm 20 \, \rm{km \, s^{-1}}$ & $416 \pm 25 \, \rm{km \, s^{-1}}$ \\\hline 
		\end{tabular}
	\end{table}
	
	We have shown that at a distance $D \approx 10 - 13 \, \rm{Mpc}$ the peculiar velocity wrt. the CMB of the observed DF2 would be unusually high. But such high peculiar velocities do arise in the observed Universe.
	For example, the galaxies NGC 1400 and NGC 1407 have a difference in velocity relative to the Hubble flow of about $1100 - 1200 \, \rm{km \, s^{-1}}$ \citep{Spolaor_2008,Tully_2015} and are probably within at least $0.1 \, \rm{Mpc}$ of each other but the relative distance uncertainties are about $4 \, \rm{Mpc}$ \citep[see table~2 in][]{Spolaor_2008}. \citet{Pawlowski_2014a} reported that the 
	NGC 3109 association forms a thin planar structure consisting of the galaxies Antila, NGC 3109, Sextans A, Sextans B, Leo P and suggested that the velocity with which this plane (which is parallel to one of the LG non-satellite galaxy planes situated symmetrically about the Milky Way-Andromeda line as discovered by \citealt{Pawlowski_2013}) is moving away from the LG is too high for a $\Lambda$CDM cosmology (the backsplash problem). Detailed modelling of the LG has shown that such a high velocity of a planar structure is indeed in conflict with the standard model of cosmology \citep{Banik_2017, Peebles_2017}. On larger scales high velocities are also observed, for example in massive interacting galaxy clusters. The two dark matter peaks of the Bullet Cluster (1E0657-65) have a relative velocity of $\approx 3000 \, \rm{km \, s^{-1}}$ \citep{Kraljic_2015}. Using FLASH-based $N$-body/hydrodynamical models \citet{Molnar_2015} demonstrated that a pre-infall relative velocity of $2250 \, \rm{km \, s^{-1}}$ is required to reproduce the observational parameters of the galaxy cluster El Gordo (ACT-CT J0102-4915).  
	The tension with standard $\Lambda$CDM cosmology arises in that an initially homogeneous and isotropic Universe needs to produce bound structures which attain such high observed relative velocities within a Hubble time. This might be alleviated in MOND cosmology, where the impact velocity of the Bullet Cluster is not problematic \citep{Angus_2008}.
	
\begin{figure}
		\centering
		\includegraphics[width=\linewidth]{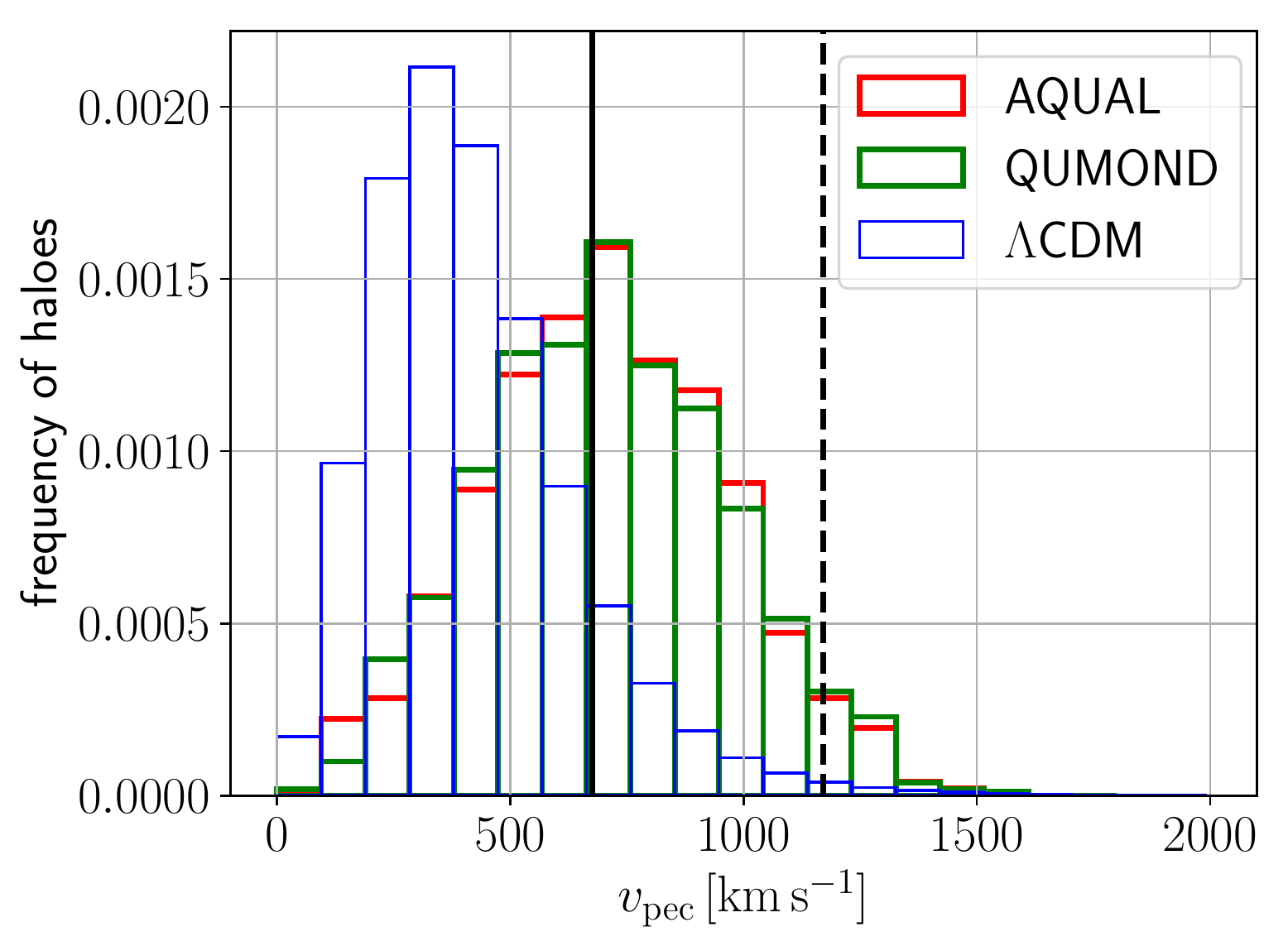}
		\caption{Peculiar velocity distribution of halos in $\Lambda$CDM (TNG300-1, $L = 302.6 \, \rm{cMpc}$) and MOND cosmologies ($L = 45.7 \, \rm{cMpc}$) using its traditional formulation as AQUAL \citep{Bekenstein_1984} and the more computer-friendly QUMOND \citep{Milgrom_2010}. The distribution of the TNG300-1 velocity field is similar to the Newtonian case shown in \citet{Candlish_2016}. The solid vertical line at $v_{\mathrm{pec}} = 676 \, \rm{km \, s^{-1}}$ marks the smallest possible peculiar velocity and the dashed line with $\sqrt{3} \, \times$ as much ($v_{\mathrm{pec}} = 1171 \, \rm{km \, s^{-1}}$) highlights the most likely peculiar velocity of the observed DF2 dwarf galaxy if located at a distance of $13 \, \rm{Mpc}$ from Earth. In the TNG300-1 simulation $1.9 \, \%$ of all subhaloes and $0.81 \, \%$ of all halos have $v_{\mathrm{pec}} \geq 1171 \, \rm{km \, s^{-1}}$. In AQUAL and QUMOND $4.2 \, \%$ and $4.7 \, \%$ of the halos have $v_{\mathrm{pec}} \geq 1171 \, \rm{km \, s^{-1}}$, respectively. The data for the MONDian cosmology are taken from \citet{Candlish_2016} (see right panel of Fig.~14).}
		\label{fig:velocity_field}
	\end{figure}

	Interestingly, if DF2 is located at $D = 13 \, \rm{Mpc}$, its resulting high peculiar velocity arises rather naturally in a MOND cosmology in which the law of gravity becomes non-Newtonian for accelerations $\la 10^{-10} \, \rm{m \, s^{-2}}$ \citep{Milgrom_1983}. The velocity field of halos is shown in Fig.~\ref{fig:velocity_field} for the TNG300-1 $\Lambda$CDM and cosmological MOND simulations conducted by \citet{Candlish_2016}. We note that the distributions of the peculiar velocity of subhaloes and halos are very similar in $\Lambda$CDM simulations such that we assume here that the initial velocity fields of the halos and subhaloes in both the $\Lambda$CDM and MOND simulations are basically the same. 
	In MOND cosmology the distribution of the present-day velocity field is shifted to higher values peaking around $700 \, \rm{km \, s^{-1}}$ compared to that obtained from standard cosmology, which peaks at around $250 \, \rm{km \,s^{-1}}$. Remarkable is that the peculiar velocity of the LG \citep[$v_{\mathrm{pec}} = 627 \pm 22 \, \rm{km \, s^{-1}}$,][]{Kogut_1993} is well consistent with the position of the peak of the MONDian velocity distribution. At $D = 13 \, \rm{Mpc}$, the most likely peculiar velocity of DF2 is $v_{\mathrm{pec}} = 1171 \,\rm{km \, s^{-1}}$ and is likely in the case of MOND. In particular, about $1.9 \, \%$ of all subhaloes in the TNG300-1 simulation have $v_{\mathrm{pec}} \geq 1171 \, \rm{km \, s^{-1}}$. In $\Lambda$CDM simulations the tails of the velocity distributions of subhaloes and halos are slightly different, because subhaloes can be boosted much easier to higher velocities. Thus, only $0.81 \, \%$ of all halos have $v_{\mathrm{pec}} \geq 1171 \, \rm{km \, s^{-1}}$ in the TNG300-1 simulation. In the two MOND versions AQUAL \citep{Bekenstein_1984} and QUMOND \citep{Milgrom_1983} about $4.2 \, \%$ and $4.7 \, \%$ of the halos have $v_{\mathrm{pec}} \geq 1171 \, \rm{km \, s^{-1}}$, respectively. The MONDian simulations are carried out in a co-moving box size of only $L \approx 46 \, \rm{cMpc}$ \citep{Candlish_2016} and thus their velocity fields are not really representative of a ``real'' MONDian Universe due to a lack of large-scale waves/fluctuations. For reasons not related to technical problems, no self-consistent hydrodynamical cosmological simulation exists, which would allow a more detailed and quantitative comparison of the structural parameters of MONDian dwarf galaxies with DF2. 
	
	In the case of distances $D \leq 13 \, \rm{Mpc}$, a further challenge for $\Lambda$CDM is the large radial peculiar velocity of DF2 relative to the motion of the LG with a large angle between their velocity vectors at the same time (see also Table~\ref{tab:LG_statistics} in Section~\ref{sec:Results_cosmological_simulations}). Given that most of the peculiar velocity of the LG arises from the GA, one would expect that the velocity vectors of two such close and fast-moving galaxies like DF2 and the LG would point at least roughly in the same direction. Indeed, the combination of a large peculiar velocity and angle between their velocity vectors reduces the detection of LG-DF2-like pairs in cosmological simulations at lower distances from the Earth.

	\subsection{Comparison with observations}
	\subsubsection{Observational constraints on the distance of DF2}
	
	At a distance of $D = 20 \, \rm{Mpc}$, DF2 and possibly also DF4 have an unusual GC population \citep{vDokkum_2018a,vDokkum_2019}. A detailed comparison with the GC population of the MW with that of DF2 shows that the tension can be solved if DF2 is located at a much smaller distance of $D \leq 13 \, \rm{Mpc}$, which is in agreement with the findings by \citet{Trujillo_2019} (see also Sections~\ref{sec:Results_Luminosities} and \ref{sec:Results_Radii}). 
	
	In contrast, \citet{vDokkum_2018c} argue that based on the radial peculiar velocity, projected sky-position, and SBF method DF2 has to be located at $D = 19.0 \pm 1.7 \, \rm{Mpc}$ and due to the non-detection of red giant branch stars one can rule out a distance of $D = 13 \, \rm{Mpc}$. However, \citet{Trujillo_2019} were able to identify the TRGB in the HST data from which they derived a distance of $D= 13.4 \pm 1.1 \, \rm{Mpc}$. In the framework of the IGIMF theory \citep{Kroupa_2003, Jerabkova_2018} the SBF method would predict a lower distance compared to \citet{vDokkum_2018c}. This is because the lower SFR of a dwarf galaxy results in a top-light IMF compared to more massive galaxies as quantified by \citet{Zonoozi_2019}. 
	
	\subsubsection{Comparison with other dwarf galaxies}
	
\begin{figure*}
		\centering
		\includegraphics[width=90mm,trim={0.5cm 0.8cm 1.0cm 0.0cm},clip]{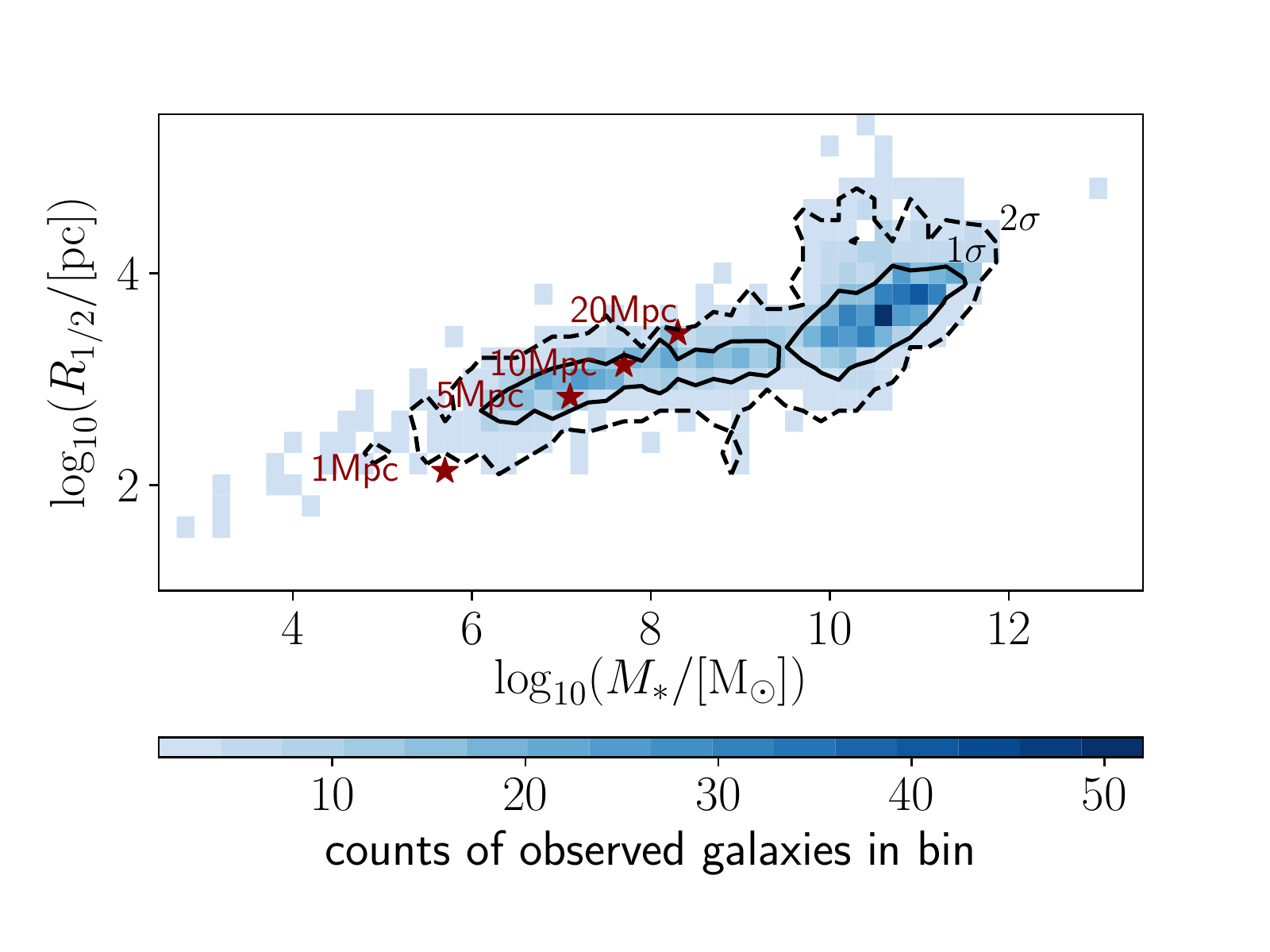}
		\includegraphics[width=85mm]{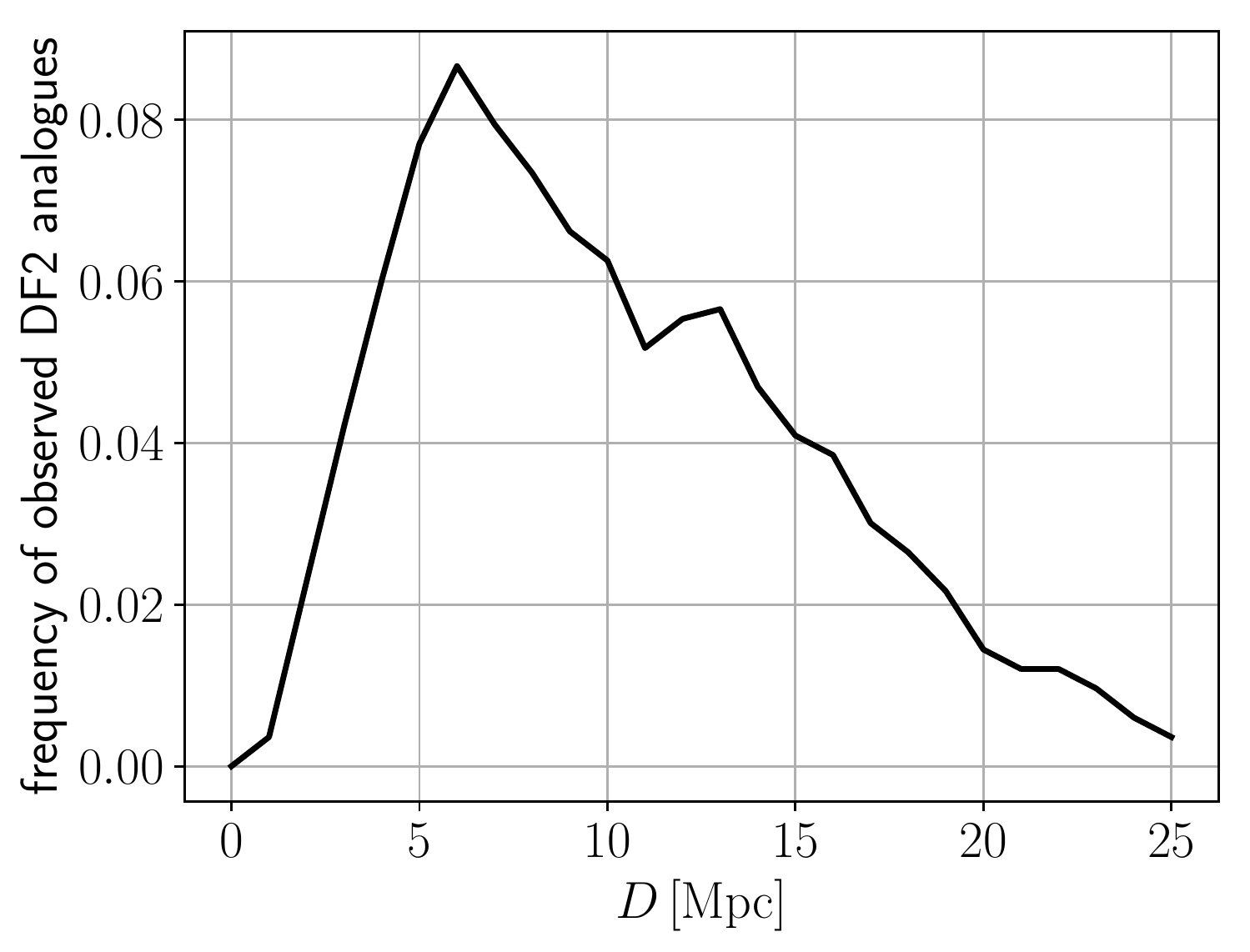}
		\caption{Left: Radius--mass diagram for observed early-type galaxies taken from \citet{Dabringhausen_2016}. The solid and dashed lines mark the $1 \sigma$ and $2 \sigma$ confidence levels, respectively. The half-light radii of the early-type galaxies are converted to 3D deprojected half-light radii by multiplying them by a factor of $4/3$ \citep[see Appendix~B in][]{Wolf_2010}. Red stars show the position of DF2 scaled for different distances. DF2 is consistent with observed dwarf galaxies with $M_{*} < 10^{9} \,  \rm{M_{\odot}}$ at the $(11 \pm 3) \, \%$ confidence level if located at $D = 20 \, \rm{Mpc}$.
		Right: Distance-dependent probability density of observed DF2 analogues in the sample of \citet{Dabringhausen_2016} with $M_{*} < 10^{9} \, \rm{M_{\odot}}$ calculated by extracting all bin values along the locus of possible DF2 positions in the radius--mass diagram (highlighted by red stars in the left-hand panel).}
		\label{fig:observations_DF2}
	\end{figure*}

	We compare in Fig.~\ref{fig:observations_DF2} (left-hand panel) the position of DF2 with observed early-type galaxies from \citet{Dabringhausen_2016} in the radius--mass plane. Restricting the sample to galaxies with $M_{*} < 10^{9} \, \rm{M_{\odot}}$ gives that DF2 is consistent with observations at a $(11 \pm 3) \, \%$ confidence level if located at $D = 20 \, \rm{Mpc}$. The radius--mass diagram also shows the position of DF2 scaled for different distances (red stars). Extracting the probability density of observed galaxies with $M_{*} < 10^{9} \, \rm{M_{\odot}}$ along this locus (right-hand panel of Fig.~\ref{fig:observations_DF2}) underpins that DF2 is, based on its size and stellar mass, not an exception compared to other observed galaxies and suggests that DF2 is statistically most likely at a distance of $D = 6 \, \rm{Mpc}$. Note that the sample of \citet{Dabringhausen_2016} also includes a substantial number of galaxies with a surface brightness lower than that of DF2.

	\subsubsection{TDGs in MOND cosmology}
	
	As already stated in Section~\ref{sec:Structural properties of subhaloes}, dark-matter-lacking galaxies are rare in current self-consistent hydrodynamical cosmological simulations. \citet{Haslbauer_2019} reported that only $0.17 \, \%$ of all galaxies with $10^{7} <M_{*}/\rm{M_{\odot}}< 10^{9}$ are dark-matter-free tidal dwarf galaxies in the Illustris-1 simulation. This is consistent with the work of \citet{Mueller_2018} in which they showed that finding phase-space correlated systems like around the Centaurus A galaxy is $ \la 0.5 \, \%$ in Illustris-1. The combined probability of simulated disc of satellites comparable to those of the Milky Way, Andromeda M31, and Centaurus A galaxy is about $5.1 \times 10^{-9}$ relative to the Millennium II and Illustris-1 simulations, when considering that $3$D information on the position of satellites is only available in these three systems \citep{Pawlowski_2014,Mueller_2018}. However, \citet{Ibata_2014} found significant evidence for such systems to be ubiquitous. Alone the existence of phase-space correlated satellite galaxies is thus in extremely significant disagreement with $\Lambda$CDM theory \citep{Kroupa_2005}. 
	
	It is expected that the number of TDGs is higher in a MONDian compared to a $\Lambda$CDM Universe.  Due to stronger self-gravitation in MONDian dynamics, TDGs can be formed more easily and be longer lasting since they are more resilient against tides from their host galaxies. Indeed, \citet{Renaud_2016} showed that Milgromian models of galaxy-encounters produce more substructures within the tidal tails. In fact, \citet{Okazaki_2000} concluded that a large fraction of observed dwarf ellipticals can be old TDGs in a cosmology where $\approx 1 - 2$ TDGs form in each galaxy--galaxy encounter and subsequently survive for a Hubble time. This seems unlikely in a $\Lambda$CDM framework because TDGs formed in one interaction are likely to be destroyed in a subsequent interaction. This is due to the higher merger rate of galaxies caused by dynamical friction between dark matter halos.

	\subsubsection{Estimated tension with $\Lambda$CDM in light of DF4} \label{sec:Estimated_tension}
	
	We have shown that the DF2 dwarf galaxy is unexpected in standard cosmology. Taking into account the dynamical $M/L$ value \citep{Martin_2018}, the stellar mass, half-light radius, GC population, peculiar velocity, and its angle relative to that of the LG, the probability of finding DF2-like galaxies in a $\Lambda$CDM cosmology is $4.8 \times 10^{-7}$ for a distance of $20.0 \, \rm{Mpc}$ (see Table~\ref{tab:results_combined} in Section~\ref{sec:combined_probability}). This rises to a maximum of $1.0 \times 10^{-4}$ for a distance of $11.5 \, \rm{Mpc}$ based on TNG100-1. Estimating that accurate mass, distance, and velocity dispersion measurements exist for 100 galaxies, there is a $\la 10^{-2}$ ($2.6 \sigma$ tension) chance of having detected a DF2-like object if at $D = 11.5 \, \rm{Mpc}$ and $ \la 4.8 \times 10^{-5}$ ($4.1 \sigma$ tension) if at $D = 20.0 \, \rm{Mpc}$. Note for completeness that adopting the \citet{vDokkum_2018a} $M/L$ value instead, the chance of detecting a DF2 like object if at $D=11.0 \, \rm{Mpc}$ is $2.4 \times 10^{-3}$ ($3.0 \sigma$ tension) and $3.6 \times 10^{-6}$ ($4.6 \sigma$ tension) if at $D=20.0 \, \rm{Mpc}$. At the lower distance, DF2 would be marginally consistent with the $\Lambda$CDM paradigm if it is the only galaxy of its kind.
	
	However, preliminary results suggest that DF4 also lacks dark matter and has similar properties to DF2 but with a smaller radial velocity and half-light radius \citep{vDokkum_2019}. The probability of DF4-like subhaloes in cosmological simulations without the constraint on its GC population is $ \leq 1.4 \times 10^{-4}$ for a distance of $20.0 \, \rm{Mpc}$, which is comparable with our estimated occurrence rate of DF2 analogues (see Table~\ref{tab:results_20Mpc} in Section~\ref{sec:Results_cosmological_simulations}). This is a conservative approach because the GCs appear to be unusually bright if located at $D = 20 \, \rm{Mpc}$ (i.e. all seven identified GCs of DF4 have absolute magnitudes $M_{\mathrm{V, 606}} \leq -8.6 \, \rm{mag}$, \citealt{vDokkum_2019}).
	
	Thus, it is very unlikely that we have observed two such dark-matter-lacking galaxies with similar properties. Assuming again measurements for $100$ galaxies the observations of DF2 and DF4 cause a $5.8 \sigma$ tension with $\Lambda$CDM theory if both are located at $D = 20.0 \, \rm{Mpc}$. Adopting $D = 11.5 \, \rm{Mpc}$ for DF2 and $D = 20.0 \, \rm{Mpc}$ for DF4, the tension becomes $4.8 \sigma$. These estimates hold true assuming the $M/L$ value by \citet{Martin_2018}. The $M/L$ value by \citet{vDokkum_2018a} would instead lead to a $6.2 \sigma$ tension with $\Lambda$CDM if both galaxies lie at $D = 20.0 \, \rm{Mpc}$, while the tension would be $5.1 \sigma$ if DF2 lies at $11.0 \, \rm{Mpc}$ and DF4 at $D = 20.0 \, \rm{Mpc}$. The \citet{vDokkum_2018a} $M/L$ value would thus virtually falsify $\Lambda$CDM cosmology. DF4 has a lower radial velocity than DF2, so there is less of an issue with making its GCs normally bright by putting it at a lower distance. However, DF4 would be then much more compact and thus more problematic for $\Lambda$CDM (Fig. \ref{fig:structure1_simulated_DF2}). A more detailed analysis of DF4 is left to a future investigation once more observational data on this galaxy become available.

	\section{Conclusion} \label{sec:Conclusion}
	The inferred lack of dark matter in the UDG DF2 by \citet{vDokkum_2018a} follows from a low intrinsic velocity dispersion of $\sigma_{\mathrm{intr}} < 10.5 \,\rm{km \, s^{-1}}$ with $90 \, \%$ confidence derived from the motion of $10$ unusual bright GCs. However, their conclusion of the discovery of a dark-matter-deficient galaxy relies on a distance of $D=20 \, \rm{Mpc}$ from Earth. DF2 and also DF4 may be satellites of NGC 1052 and the whole system may be located at a much shorter distance of $\approx 10 \, \rm{Mpc}$, making their GC populations comparable to those of other galaxies \citep{Haghi_2019}. In the here presented analysis, we showed that finding similar galaxies in standard cosmology is extremely rare regardless of its actual distance from Earth. We summarize our results as follows:
	
	Constraints on the stellar half-mass, stellar mass, and total-to-stellar mass ratio consistent with the physical description of \citet{Martin_2018} cause a decrease of the frequency of DF2-like subhaloes in $\Lambda$CDM simulations by about two orders of magnitude. These structural properties of simulated subhaloes mildly prefer a distance of $D = 11.5_{-1.5}^{+4.0} \, \rm{Mpc}$ for the TNG100-1 simulation in which the structural probabilities become maximal. 
	
	However, at $D \leq 13 \, \rm{Mpc}$, DF2 has a high peculiar velocity wrt. the CMB of $v_{\mathrm{pec}} \geq 676 \, \rm{km \, s^{-1}}$, which differs from that of the LG by $v_{\mathrm{rel}} \geq 1112 \, \rm{km \, s^{-1}}$ with a large angle of $\theta \approx 117 \, ^{\circ}$. Taking into account only the peculiar velocity wrt. the CMB and motion relative to LG-like subhaloes yields a maximum probability at $D = 20.0 \, \rm{Mpc}$ (the limit of the prior) with a $1\sigma$ ($2\sigma$) lower limit of $16.5 \, \rm{Mpc}$ ($12.0 \, \rm{Mpc}$). In particular, these peculiar velocity criteria alone decrease the frequency of DF2-like subhaloes by two orders of magnitude at $D \la 13 \, \rm{Mpc}$ and by about one order of magnitude at $D \ga 16 \, \rm{Mpc}$. This distance-dependence of the peculiar velocity probability is caused by the peculiar velocity of DF2 wrt. the CMB reference frame rather than by its motion relative to the LG. Interestingly, the peculiar velocity probability is the same for TNG100-1 and the much larger TNG300-1 simulation in which longer wave modes and therefore more massive galaxy clusters are included. We demonstrated that the velocity field of TNG300-1 has converged by comparing it with that of the Millennium simulation, which has about a two times larger box length than TNG300-1 (Section \ref{sec:Velocity field of subhaloes}). Since the reduction of the frequency based on the peculiar velocity of DF2-like subhaloes is the same for both the TNG100-1 and TNG300-1 simulations and the fact that TNG100-1 has a much better resolution than TNG300-1, our main results rely on the former simulation run. 
	
	Such a high velocity, which arises if DF2 is at $D \la 13 \, \rm{Mpc}$, is not impossible in $\Lambda$CDM cosmology, but reduces the detection probability of simulated DF2 analogues for lower distances significantly. It is interesting to point out that high peculiar velocities can be occasionally observed in the real Universe and are likely in Milgromian gravitation, but not in $\Lambda$CDM cosmology (Section~\ref{sec:Velocity field of subhaloes} in this work and Fig.~14 in \citealt{Candlish_2016}). Thus, the detection probability of DF2 analogues located at $D \leq 13 \, \rm{Mpc}$ is higher in a MOND cosmology, because the velocity field would be shifted to higher values compared to that of the $\Lambda$CDM model. However, so far and for a variety of reasons not related to technical issues, there are no self-consistent hydrodynamical MOND cosmological simulations available which would allow a detailed comparison of the structural properties, masses, and kinematics of MONDian dwarf galaxies with observations.
	
	Considering the structural properties and peculiar velocity of subhaloes, cosmological $\Lambda$CDM simulations suggest that DF2 analogues occur with a probability of $1.2 \times 10^{-3}$ for a distance of $20.0 \, \rm{Mpc}$. The $1\sigma$ ($2\sigma$) lower distance limit is $16.0 \, \rm{Mpc}$ ($11.5 \, \rm{Mpc}$) based on TNG100-1 (Section~\ref{sec:Results_cosmological_simulations}).

	The observed DF2 galaxy and presumably also DF4 have an extremely unusual GC population if located at $D = 20 \, \rm{Mpc}$. Bringing DF2 to a much smaller distance of about $D = 10 -13 \, \rm{Mpc}$ makes the luminosities and structural properties of its GC population comparable with that in other observed galaxies (Sections~\ref{sec:Results_Luminosities} and \ref{sec:Results_Radii}) as also concluded by \citet{Trujillo_2019}. Since GCs are in the high-acceleration regime of MOND, the GC population of DF2 is independent of the cosmological model and would therefore challenge any cosmological model if at $D = 20 \, \rm{Mpc}$. 
	
	Combining the results from the cosmological $\Lambda$CDM simulation with the luminosity of the GC population, DF2 is statistically expected at a distance of $D = 11.5\pm1.5 \, \rm{Mpc}$ with a maximal probability of $1.0 \times 10^{-4}$ based on TNG100-1 (Section~\ref{sec:combined_probability}). Restricting this analysis only for DF2 analogues embedded in host halos with $M_{200} \leq 10^{13} \, \rm{M_{\odot}}$ reduces the probability again by about one order of magnitude and yields a most likely distance of $D = 13.0\pm1.5 \, \rm{Mpc}$ (Appendix~\ref{appendix:combined_probability}). Thus, our results are consistent with \citet{Trujillo_2019}. This is in disagreement with the results of \citet{vDokkum_2018c} who argued that DF2 cannot be located at $D \leq 13 \, \rm{Mpc}$. However, \citet{Trujillo_2019} derived a distance of $13.0 \pm 0.4 \, \rm{Mpc}$ based on redshift-independent indicators. 	
	If DF2 is located at $20.0 \, \rm{Mpc}$ as proposed by \citet{vDokkum_2018a}, its combined probability of occurrence in standard cosmology is $\leq 4.8 \times 10^{-7}$ (Section~\ref{sec:Estimated_tension}). 
	
	The chance of detecting DF4-like subhaloes in cosmological simulations without the constrain on its GC population is $ \leq 1.4 \times 10^{-4}$ for a distance of $20.0 \, \rm{Mpc}$ (Section \ref{sec:Results_cosmological_simulations}). We have not studied the GC population and frequency of DF4-like objects over distance, because precise measurements of its GC population are not publicly available yet. 
	
	Taking into account that precise measurements exist for $100$ galaxies, DF2 is in tension with standard cosmology at $2.6 \sigma$ ($4.1 \sigma$) if located at $11.5 \, \rm{Mpc}$ ($20.0 \, \rm{Mpc}$). 
	Adopting the former distance for DF2 and ignoring the GC population of DF4, both dark matter lacking galaxies would cause a $4.8 \sigma$ tension. Placing both galaxies at $D = 20.0 \, \rm{Mpc}$ results in a rejection of the $\Lambda$CDM cosmology by $5.8 \sigma$. If the recently observed DF4 dwarf galaxy \citep{vDokkum_2019} has a similar GC population to DF2, the tension with $\Lambda$CDM becomes even more significant (Section~\ref{sec:Estimated_tension}). Therefore, the observed properties of dwarf galaxies may point to the need for new physics. 
	
	\section*{Acknowledgements}
	We thank the anonymous referee for her/his suggested improvements. IB is supported by an Alexander von Humboldt postdoctoral research fellowship. KG was supported by the German-Russian Interdisciplinary Science Center funded by the German Federal Foreign Office via the German Academic Exchange Service. We thank Graeme Candlish for sharing his data on the velocity field in MOND cosmology and Hosein Haghi for useful discussions about the properties of NGC 1052-DF2. 
	
	\bibliographystyle{mnras}
	\bibliography{ref}
	
	\begin{appendix}
		
	\section{Scaling observables to different distances} \label{appendix:scaling_relations}
	
	In order to scale the properties to different distances, $D$, we apply in Sections~\ref{sec:Methods_selection_criteria_algorithm}, \ref{sec:Results}, and \ref{sec:Discussion} the following scaling relations for the stellar mass, $M_{*}$, stellar half-mass radius, $R_{1/2}$, total-to-stellar ratio, $M_{\mathrm{total}}/M_{*}$, and luminosity, $L$:
	
	\begin{eqnarray}
		M_{*} &\propto& D^{2}, \\
		R_{1/2} &\propto& D , \\
		M_{\mathrm{total}}/M_{*} &\propto& D^{-1}, \\
		L &\propto& D^{2}.
		\label{eq:scaling_relations}
	\end{eqnarray}
	
	\section{Combined Probabilities} \label{appendix:combined_probability}
\begin{figure}
		\centering
		\includegraphics[width=\linewidth]{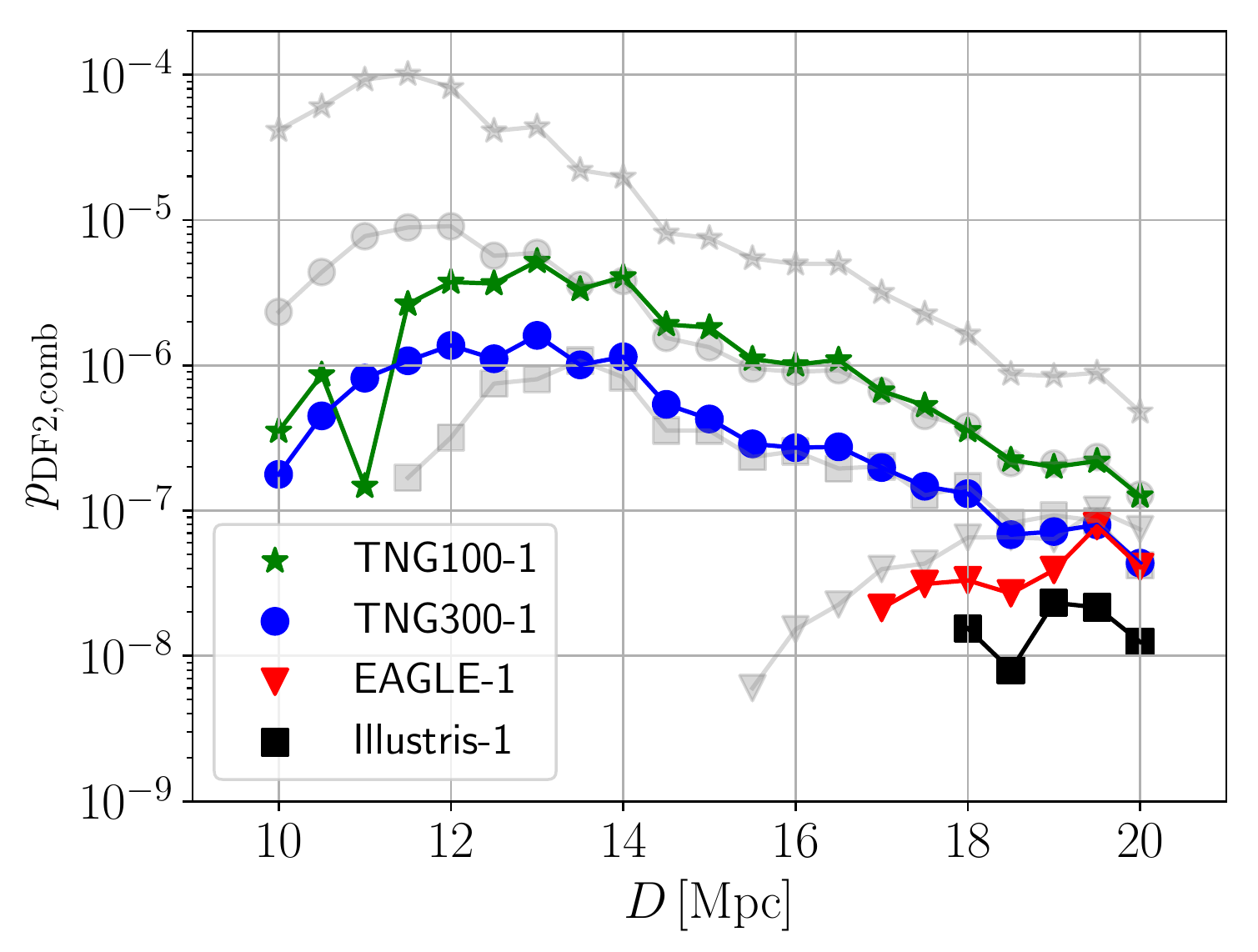}
		\caption{Same as Fig.~\ref{fig:Pcomb2} (grey data points) but only for DF2-like subhaloes embedded in host halos with $M_{200} \leq 10^{13} \, \rm{M_{\odot}}$. There are no DF2-like subhaloes in the EAGLE-2 and EAGLE-3 simulations.}
		\label{fig:Pcomb2_M200}
	\end{figure}

	DF2 and the massive elliptical galaxy NGC 1052 have a statistically expected 3D separation of about $100 \, \rm{kpc}$ if located at $20 \, \rm{Mpc}$ \citep{vDokkum_2018a}. Thus, within the $\Lambda$CDM framework DF2 can be embedded in the main halo of NGC 1052, which has a mass of $M_{200} \approx 10^{13} \, \rm{M_{\odot}}$  estimated from the stellar mass of NGC 1052 ($M_{*} \approx 10^{11} \, \rm{M_{\odot}}$). Therefore, only DF2-like subhaloes embedded in host halos with  $M_{200} \leq 10^{13} \, \rm{M_{\odot}}$ are included in Fig.~\ref{fig:Pcomb2_M200}. As in Section~\ref{sec:combined_probability} the probability is maximal in the TNG100-1 simulation and DF2 is most likely at $D = 13.0 \pm 1.5 \, \rm{Mpc}$ with a probability of $5.2 \times 10^{-6}$ ($4.6 \sigma$). For the Illustris-1, TNG100-1, and TNG300-1 the probability is reduced by a factor of $\approx 10$. The results of the EAGLE-1 simulation are not much affected, because here the probability is already quite low. No DF2 analogues can be found for $D < 17 \, \rm{Mpc}$ and $D < 18 \, \rm{Mpc}$ in the EAGLE-1 and Illustris-1 simulations, respectively.
	
	\end{appendix}
	
	\bsp
	\label{lastpage}
	
\end{document}